\documentclass[12pt]{article}
\usepackage{latexsym}
\usepackage[mathscr]{eucal}
\usepackage{amsfonts}
\usepackage{amscd}
\usepackage{cite}
\usepackage{amssymb}
\usepackage[centertags]{amsmath}
\usepackage{dsfont}
\usepackage{enumerate}
\usepackage{graphicx}
\usepackage{float}
\usepackage{epsfig}

\usepackage{multirow}
\usepackage{amsmath}

\usepackage{array}
\newcolumntype{P}[1]{>{\centering\arraybackslash}p{#1}}
\newcolumntype{M}[1]{>{\centering\arraybackslash}m{#1}}

\restylefloat{figure}

\usepackage{array} 
\usepackage{amssymb}
\usepackage{graphics,graphpap}
\usepackage{graphicx}
\usepackage{color}
\usepackage{graphicx}
\usepackage{dcolumn}
\usepackage{epsfig}
\usepackage{epstopdf}
\usepackage{bm}
\usepackage{amsmath}
\usepackage{amsfonts}
\usepackage{textcomp}
\usepackage{bbm}
\usepackage{setspace}
\usepackage[centertags]{amsmath}

\usepackage[usenames,dvipsnames]{xcolor}

\def\simge{\mathrel{%
   \rlap{\raise 0.511ex \hbox{$>$}}{\lower 0.511ex \hbox{$\sim$}}}}

\def\simle{\mathrel{
   \rlap{\raise 0.511ex \hbox{$<$}}{\lower 0.511ex \hbox{$\sim$}}}}

\def\s#1{\setbox0=\hbox{$#1$}%
\rlap{\ifdim\wd0>.7em\kern.22\wd0\else\kern.1\wd0\fi /}#1}

\newcommand{\cleqn}{\setcounter{equation}{0}}

\newcommand{\newc}{\newcommand}
\newc{\be}{\begin{equation}}
\newc{\ee}{\end{equation}}
\newc{\bea}{\begin{eqnarray}}
\newc{\eea}{\end{eqnarray}}
\newc{\ben}{\begin{equation*}}
\newc{\een}{\end{equation*}}
\newc{\bean}{\begin{eqnarray*}}
\newc{\eean}{\end{eqnarray*}}
\newc{\ol}{\overline}
\newc{\wt}{\widetilde}
\newc{\bs}{\boldsymbol}
\newc{\m}{\mathcal}
\newc{\la}{\lambda}
\newc{\lra}{\longrightarrow}
\newc{\vp}{\varphi}
\newc{\ti}{\tilde}
\newcommand{\vev}[1]{\langle#1\rangle}

\begin{document}

\title{\hfill ~\\[-40mm]
          \hfill\mbox{\small  SI-HEP-2015-24}\\[-3.5mm]
          \hfill\mbox{\small  QFET-2015-30}\\[13mm]
      \textbf{Approaching Minimal Flavour Violation from an 
      $\boldsymbol{SU(5)\times S_4\times U(1)}$ SUSY GUT}}
\date{}

\author{\\Maria Dimou\footnote{E-mail: {\tt md1e10@soton.ac.uk}}$\;\;^a$, Stephen F. King\footnote{E-mail: {\tt king@soton.ac.uk}}$\;\;^a$,
Christoph Luhn\footnote{E-mail: {\tt  christoph.luhn@uni-siegen.de}}$\;\;^b$\\[10mm]
 $^a$ \emph{\small{}School of Physics and Astronomy, University of Southampton,}\\
  \emph{\small Southampton, SO17 1BJ, United Kingdom}\\[2mm]
 $^b$ \emph{\small Theoretische Physik 1, Naturwissenschaftlich-Technische Fakult\"at,}\\
  \emph{\small Universit\"at Siegen, Walter-Flex-Stra{\ss}e 3, 57068 Siegen, Germany}}

\maketitle

\begin{abstract}
\noindent 
We show how approximate Minimal Flavour Violation (MFV) can emerge from 
an $SU(5)$ Supersymmetric Grand Unified Theory (SUSY GUT) supplemented by an
$S_4 \times U(1)$ family symmetry, which provides a good description of all
quark and lepton (including neutrino) masses, mixings and CP violation. 
Assuming a SUSY breaking mechanism which respects the family symmetry, 
we calculate in full explicit detail the low energy mass insertion parameters
in the super-CKM basis, including the effects of canonical normalisation and
renormalisation group running. We find that the very simple family symmetry
$S_4 \times U(1)$ is sufficient to approximately reproduce the effects of low
energy MFV.
\end{abstract}
\thispagestyle{empty}
\vfill

\newpage


\thispagestyle{empty}

\tableofcontents

\vspace{5mm}

\hrulefill



\setcounter{page}{1}

\section{Introduction}
\cleqn
The mystery of flavour has been with us from the discovery of the muon in 1936
to the discovery of neutrino mass and mixing in 1998. The Standard Model (SM),
extended to include neutrino mass, is described by at least 26 parameters, of
which no less than 20 are flavour parameters: 
10 from the quark sector and at least 10 from the lepton sector. At least two
of these parameters are related to CP violation in the quark and lepton
sectors, although the latter has not yet been definitively observed. 

A lot of effort has been put into trying to understand the flavour structure
of the~SM
(for reviews see e.g.~\cite{King:2015aea}). Its peculiar features include
hierarchical charged fermion masses, with the down-type quark and charged
lepton masses showing a similar pattern which differs from that of the up-type
quarks, while neutrinos are significantly lighter than all other
particles. Flavour mixing in the lepton sector has turned out to be much
larger than in the quark sector, and the number of generations is not explained. 

Following the award of the 2015 Nobel Prize for ``the discovery of neutrino
oscillations which shows that neutrinos have mass'', we still have no more
understanding of flavour than back in 1936 when Rabi famously asked of the
muon ``who ordered that?''. Part of the reason for this impasse is the failure
of experiment to measure any flavour and CP violation beyond that expected in 
the SM. The problem is that the SM is not a theory of flavour and, as such, 
provides no understanding of the origin or nature of flavour.

In the absence of any observed beyond SM flavour and CP violation, a sort of
``straw man'' ansatz for flavour has emerged known as Minimal Flavour
Violation (MFV)~\cite{Buras:2000dm} in which all flavour and
CP-violating transitions are postulated to originate in the SM Yukawa matrices so
that they are governed by the CKM matrix. The formulation of MFV in an
effective field theory involving a high-energy $SU(3)^5$ flavour symmetry, broken
only by the Yukawa matrices, allows higher-dimensional operators which can
contribute considerably to flavour
observables~\cite{D'Ambrosio:2002ex,Bobeth:2002ch}. 
Going beyond an effective field theory description, it is possible to
implement the idea of MFV in a renormalisable theory by introducing new heavy
fermions. In such a setup, the flavour symmetry is broken by scalar fields
whose Vacuum Expectation Values (VEVs) are related to the Yukawa matrices in
an inverse way~\cite{Grinstein:2010ve}. Although this differs
from the standard MFV approach, where the fundamental flavour breaking fields
are linearly related to the Yukawa matrices, it does reproduce MFV
phenomenologically by predicting very SM-like flavour and CP violation, which
is of course exactly what is observed.

When considering extensions of the SM, such as Supersymmetry (SUSY) 
softly broken at the TeV scale, then in general large deviations from SM
flavour and CP violation are expected. SUSY models include
one-loop diagrams that lead to Flavour Changing Neutral Current (FCNC)
processes such as e.g. $b\to s\gamma$ and $\mu \to e \gamma$ at rates which
are proportional to the size of the off-diagonal elements of the scalar mass
matrices, when the latter have been rotated to the super-CKM (SCKM) basis
where the Yukawa matrices are diagonal~\cite{Chung:2003fi}. 
These SUSY contributions are tamed in the Constrained Minimal Supersymmetric
Standard Model (CMSSM) which postulates that, at the high energy scale, the
SUSY breaking squark and slepton mass squared matrices are proportional to the
unit matrix and the trilinear $A$-terms are additionally aligned with the Yukawa
matrices, resulting in an (approximate) MFV-like structure at low
energy~\cite{Chung:2003fi}.

In the framework of Grand Unified Theories (GUTs), the embedding of the SM
fermions into GUT multiplets does not allow to implement the
$SU(3)^5$ flavour symmetry of MFV. However, in GUTs based on
$SU(5)$~\cite{Georgi:1974sy} or the Pati-Salam 
group $SU(4)\times SU(2)\times SU(2)'$~\cite{Pati:1973uk}, it is possible to
introduce an $SU(3)^2$ flavour symmetry instead, and this has been shown to
lead to sufficient suppression of flavour violation~\cite{Barbieri:2015bda}.
Considering SUSY GUTs, the CMSSM framework always provides a safe haven from
unwanted flavour violation, although CP violation in the form of Electric
Dipole Moments (EDMs) remains a challenge~\cite{Chung:2003fi}. However, with
SUSY and SUSY GUTs, the real challenge is to justify the assumptions of MFV or
the CMSSM, while at the same time providing a realistic explanation of quark
and lepton (including neutrino) masses, mixing and CP violation. This
non-trivial balancing act is what concerns us in this paper.

The discovery of neutrino mass and mixing has spurred a lot of work aiming to
describe flavour in terms of a family symmetry of some kind, in particular
discrete non-Abelian family symmetry~\cite{King:2015aea}. It was realised
early on that in such models, the idea of spontaneous flavour and CP violation
could effectively tame the flavour and CP problems of the
SM~\cite{Ross:2002mr,Ross:2004qn} without any {\it ad hoc} assumptions about
MFV or the CMSSM. 
The main point is that the same family symmetry introduced to understand the
Yukawa sector will also automatically control the flavour structures of the soft
SUSY breaking sector. The only requirement is that the SUSY breaking hidden
sector must respect the family symmetry, which means that the family (and CP)
symmetry breaking scale must be below the mass scale of the messengers which
mediate SUSY breaking to the visible sector. SUSY breaking in the framework of
supergravity provides one attractive example for such a situation.

The idea of using family symmetry to solve the SUSY favour and CP problems
has been fully explored in the framework of an $SU(3)$ family
symmetry~\cite{Ross:2004qn,Antusch:2007re}, where it was shown that 
the flavons that spontaneously break family and CP symmetry will perturb the
SUSY breaking sector, leading to tell-tale signatures of flavour and CP
violation beyond MFV or the CMSSM. Unfortunately, these signatures which were
expected to appear in Run1 of the LHC~\cite{Altmannshofer:2009ne} did not in
fact materialise, and indeed the allowed parameter space has been much
reduced~\cite{Buras:2012ts}.

In the setup discussed in~\cite{Antusch:2007re}, the extra flavour violation
can be understood as follows. At leading order, the CMSSM is enforced by the
$SU(3)$ family symmetry acting on the squark and slepton mass squared matrices.
However the fact that $SU(3)$ is broken by flavons, as it must be to generate
the quark and lepton masses, means that flavons appearing in the K\"ahler
potential will give important contributions to the kinetic terms, requiring
extra canonical normalisation~\cite{King:2003xq}. Since SUSY breaking also
originates from the K\"ahler potential, the flavons will also modify the
couplings of squarks and sleptons to the fields with SUSY breaking $F$-terms. 
The resulting corrections to the soft mass squared matrices from unity will be
similar to the corrections of the corresponding K\"ahler metrics, yet both are
not aligned due to independent coefficients of the 
relevant operators. Likewise, the trilinear soft SUSY breaking $A$-terms will
replicate the flavour structure of the Yukawa matrices prior to canonical
normalisation, but exact alignment is not realised. All of this occurs at the
high scale. Additional flavour violation is generated by renormalisation group
(RG) running down to low energy, taking into account the seesaw
mechanism~\cite{seesaw} which will involve thresholds at an intermediate scale, see 
e.g.~\cite{Yamada:1992kv,Antusch:2008tf}.  

In this paper we show how approximate MFV can emerge from an $SU(5)$ SUSY GUT, 
supplemented by an $S_4 \times U(1)$ family symmetry~\cite{M1,M2},
which provides a good description of all quark and lepton (including neutrino)
masses, mixings and CP violation. Assuming that SUSY breaking respects the
family symmetry, we calculate in full detail the low energy mass insertion
parameters in the SCKM basis. We include the effects of canonical
normalisation as well as RG running. Remarkably, due to the peculiar flavour
structure of the model, we find that the small family symmetry $S_4 \times
U(1)$ is sufficient to reproduce the effects of low energy MFV much more
accurately than the previous $SU(3)$ family symmetry model.




\section{Trimaximal $\boldsymbol{S_4 \times SU(5)}$ model}
\cleqn

In this section, we present the basic ingredients of the supersymmetric
model of flavour proposed in~\cite{M2}. It is capable of
correctly describing a sizable reactor neutrino mixing angle $\theta^l_{13}$ by
generating a neutrino mass matrix of trimaximal form. The model represents a
modification of an earlier tri-bimaximal model~\cite{M1} with
only minor changes.
Being formulated in a supersymmetric $SU(5)$ grand unified framework, the
matter superfields fall into the $\bf{10}$ and $\bf{\bar{5}}$ representations,
\begin{eqnarray}
T&=&\frac{1}{\sqrt{2}}\left(\begin{array}{ccccc}
0&-u^c_G & u^c_B &-u_R &-d_R\\
u^c_G &0 &-u^c_R&-u_B&-d_B\\
-u^c_B&u^c_R&0&-u_G&-d_G\\
u_R&u_B&u_G&0&-e^c\\
d_R&d_B&d_G&e^c&0
\end{array}\right)~~\text{and}~~~F=\left(d^c_R~d^c_B~d^c_G~e~-\nu\right),
\end{eqnarray}%
where the superscript $c$ denotes charge conjugation of the right-handed
superfields. Table~\ref{particles} lists the matter, Higgs and flavon superfields
together with their transformation properties under the imposed $SU(5) \times
S_4 \times U(1)$ symmetry. Details of the non-Abelian finite group $S_4$ 
are provided in Appendix~\ref{App:S4GroupTheory}.
\begin{table}
\begin{center}
$$
\begin{array}{|c||c|c|c|c||c|c|c||c|c|c|c|c|c|c|c|c|}\hline
\text{Field}\!\!\phantom{\Big|} & ~T_3~ & ~T~ & ~F~ & ~N~ & ~H^{}_{5}~ & ~H_{\ol{5}}~ & ~H_{\ol{45}}~ &  ~\Phi^u_2~ & ~\wt\Phi^u_2~ & ~\Phi^d_3~ & ~\wt\Phi^d_3~ & ~\Phi^d_2~  & ~\Phi^\nu_{3'}~ & ~\Phi^\nu_2~ & ~\Phi^\nu_1 ~& ~\eta~\\\hline
SU(5)\!\!\phantom{\Big|} & \bf 10 & \bf 10 & \bf \ol 5 & \bf 1 &\bf  5 &\bf \ol 5 &\bf \ol{45}
&\bf 1&\bf 1&\bf 1&\bf 1&\bf 1&\bf 1&\bf 1&\bf 1& \bf 1\\\hline
S_4\!\!\phantom{\Big|} & \bf 1&\bf 2&\bf 3&\bf 3&\bf 1&\bf 1&\bf 1&\bf 2&\bf 2&\bf 3&\bf 3&\bf
2& \bf {3^{\prime}}&\bf 2&\bf 1 & \bf {1^{(\prime)}}\\\hline
U(1)\!\!\phantom{\Big|} & 0&5&4&-4&0&0&1&-10&0&-4&-11 &1 &8&8&8& 7 \\\hline
\end{array}%
$$
\end{center}
\caption{\label{particles}The matter, Higgs and flavon superfields of the
  model in~\cite{M2} together with their transformation properties under the
  imposed $SU(5)\times S_4 \times U(1)$ symmetry.}  
\end{table}
The ${\bf {\bar{5}}}$-plets, labelled by $F$, are assigned to a triplet
representation of $S_4$, while the ${\bf{10}}$-plets are split into an $S_4$
doublet $T$ for the first two generations and an $S_4$ singlet $T_3$ for the
third generation. In addition, right-handed neutrinos $N$ are introduced 
transforming in the same $S_4$ triplet representation as $F$.
The $SU(5)$ Higgs fields $H_{\bf{5}}$,  $H_{\bf{\bar{5}}}$ and $H_{\bf{\bar{45}}}$
are all $S_4$ singlets.  Note that each of these GUT Higgs representations
contains an $SU(2)_L$ Higgs doublet. Therefore, the low energy doublet $H_u$
originates from $H_{\bf{5}}$, while $H_d$ arises from a linear combination of
$H_{\bf{\bar 5}}$ and
$H_{\bf{\bar{45}}}$~\cite{Chung:2003fi,SPrimer}.\footnote{As 
$H_{\bf{\bar  5}}$ and $H_{\bf{\bar{45}}}$ transform differently under $U(1)$,
it is clear that the mechanism which spawns the low energy Higgs doublet $H_d$
must necessarily break $U(1)$. Although the discussion of any details of the
$SU(5)$ GUT symmetry breaking (which, e.g., could even have an extra
dimensional origin) are beyond the scope of our paper, we remark that
a mixing of $H_{\bf{\bar 5}}$ and  $H_{\bf{\bar{45}}}$ could be induced by 
introducing the pair $H^\pm_{\bf{24}}$ with $U(1)$ charges $\pm
1$ in addition to the standard $SU(5)$ breaking Higgs $H^0_{\bf{24}}$.} 
In addition, we introduce a number of flavon fields $\Phi^f_{\rho}$, which are
labelled by the corresponding $S_4$ representation~$\rho$ as well as the fermion
sector~$f$ to which they couple at leading order (LO). 
Two flavons, $\Phi^u_2$ and $\tilde{\Phi}^u_2$, generate the LO up-type quark
mass matrix. Three flavon multiplets, $\Phi^d_3$, $\tilde{\Phi}^d_3$ and
$\Phi^d_2$, are responsible for the down-type quark and charged lepton mass
matrices. Finally, the right-handed neutrino mass matrix is generated  from
the flavon multiplets $\Phi^\nu_{3'}$,  $\Phi^\nu_2$ and $\Phi^\nu_1$  as well as
the flavon $\eta$ which is responsible for breaking the tri-bimaximal pattern
of the neutrino mass matrix to a trimaximal one at subleading
order~\cite{M2}. The additional $U(1)$ symmetry has been introduced in order
to control the coupling of the flavon fields to the matter fields in a way
which avoids significant perturbations of the LO flavour structure by
higher-dimensional operators. We refer the reader to~\cite{M1} for more details.  

The vacuum structure of the flavon fields arises from the $F$-term alignment
mechanism~\cite{Altarelli:2005yp}. Introducing a set of
so-called driving fields, the corresponding $F$-term conditions give rise to
particular flavon alignments as described in Appendix~\ref{App:AppendixA}. To
LO, these are given as~\cite{M1, M2},
\begin{eqnarray}
\label{LeadingFlavonVevs1}
\frac{\langle \Phi^u_2 \rangle}{M}&=& \left(
\begin{array}{ccc}
0\\[-0.5mm]
1
\end{array}
\right)\phi^u_2\,\lambda^4,
\qquad
 \frac{\langle \tilde{\Phi}^u_2 \rangle}{M} = \left(
\begin{array}{ccc}
0\\[-0.5mm]
1
\end{array}
\right)\tilde{\phi}^u_2\,\lambda^4,\\[-1mm]
\label{LeadingFlavonVevs2}
 \frac{\langle \Phi^d_3 \rangle}{M}&=&\left(
\begin{array}{ccc}
0\\[-0.5mm]
1\\[-0.5mm]
0
\end{array}
\right)\phi^d_3\,\lambda^2,
\qquad
\frac{\langle \tilde{\Phi}^d_3 \rangle}{M}=\left(
\begin{array}{ccc}
0\\[-0.5mm]
-1\\[-0.5mm]
1
\end{array}
\right) \tilde{\phi}^d_3\,\lambda^3, 
\qquad
\frac{\langle \Phi^d_2 \rangle}{M}=\left(
\begin{array}{ccc}
1\\[-0.5mm]
0
\end{array}
\right)\phi^d_2\,\lambda\ ,\\[-1mm]
\label{LeadingFlavonVevs3}
\frac{\langle \Phi^\nu_{3'} \rangle}{M}&=&\left(
\begin{array}{ccc}
1\\[-0.5mm]
1\\[-0.5mm]
1
\end{array}
\right)\phi^\nu_{3'}\,\lambda^4,
\qquad
\frac{\langle \Phi^\nu_2 \rangle}{M}=\left(
\begin{array}{ccc}
1\\[-0.5mm]
1
\end{array}
\right)\phi^\nu_2\,\lambda^4,
~\quad
 \frac{\langle \Phi^\nu_1 \rangle}{M}= \phi^{\nu}_1\,\lambda^4 ,
~\quad
 \frac{\langle \eta \rangle}{M} = \phi^\eta\,\lambda^4,~~~~~~
\end{eqnarray}\\[-3.5mm]
where $\lambda\approx 0.225$ is the Wolfenstein
parameter~\cite{Wolfenstein:1983yz} and the $\phi$s are dimensionless order 
one parameters. Imposing CP symmetry of the underlying theory, all coupling
constants can be taken real~\cite{S4CP,Luhn:2013vna}, so that CP is broken
spontaneously by generally complex values for the $\phi$s.
$M$~denotes a generic messenger scale which is common to all
the non-renormalisable effective operators and assumed to be around the scale
of grand unification. Considering also subleading terms in the flavon potential,
these LO vacuum alignments receive corrections which are parameterised by
small shifts as discussed in Appendix~\ref{App:AppendixA}, and shown
explicitly in Eq.~(\ref{eq:vevs+shifts}). Throughout our calculations, we
have taken into account such shifts as well as all other subleading effects. 
As our LO results for the mass insertion parameters depend solely on the LO
structure of the model, we only report the LO analysis in the main part of
this paper. When giving explicit expressions, we therefore limit ourselves to
showing the leading contributions, omitting additional higher order
corrections. We will indicate such approximations by $\approx$ throughout the paper. 
Finally, the VEVs of the two neutral Higgses are:

\vspace{-7mm}

\begin{eqnarray}
\upsilon_u&=&\frac{\upsilon}{\sqrt{1+t_\beta^2}}t_\beta,~~~~~~\upsilon_d=\frac{\upsilon}{\sqrt{1+t_\beta^2}},
\end{eqnarray}\\[-3mm]
where $t_\beta\equiv \tan(\beta)=\frac{\upsilon_u}{\upsilon_d}$
and $\upsilon = \sqrt{\upsilon_u^2+\upsilon_d^2}=174$ GeV.




\section{K\"{a}hler potential}
\cleqn

A characteristic feature of any effective theory is the presence of
non-renormalisable operators which are only constrained by  the imposed
symmetries. In the context of supersymmetry, this is the case for both the 
superpotential as well as the K\"ahler potential. The effective coupling of flavon
fields to the K\"ahler potential gives rise to kinetic terms with a
non-canonical K\"ahler metric $\mathcal K \neq \mathds 1$,

\vspace{-2.5mm}

\begin{equation}
\mathcal{L}_{\text{kin}}~=~\mathcal{K}_{ij}\left(\partial_\mu \tilde
f^*_i\partial^\mu \tilde f_j
+i \, f^*_i\partial_\mu \bar{\sigma}^\mu f_j\right),
\end{equation}\\[-4mm]
where $\tilde f$ and $f$ are, respectively, the scalar and fermionic
components of a generic chiral superfield $\hat f$. In order to extract
physically meaningful properties of a model, the kinetic terms have to be
brought to a canonical form. The required basis transformation is usually
referred to as canonical normalisation~\cite{King:2003xq}. 

In the context of $SU(5)$, we encounter a K\"ahler metric for each of the three 
GUT representations containing the matter fields. We denote these by $\mathcal K_T$,
$\mathcal K_F$ and $\mathcal K_N$, respectively. Using the symmetries of
Table~\ref{particles}, the expansions of these $3 \times 3$ matrices in terms
of flavon fields can be obtained from 
\begin{eqnarray}
\begin{pmatrix} T^\dagger &  T^\dagger_3 \end{pmatrix} 
(\mathcal K_T - \mathds 1)  
 \begin{pmatrix} T \\ T_3 \end{pmatrix}&=& \sum_n~
\begin{pmatrix} T^\dagger &  T^\dagger_3 \end{pmatrix} 
 \begin{pmatrix}
c^{K_{T_{22}}}_{n} \left(\mathcal{R}_2\right)_n & 
c^{K_{T_{i3}}}_{n} \left(\mathcal{R}_4\right)_n \\
\left[c^{K_{T_{i3}}}_{n} \left(\mathcal{R}_4\right)_n\right]^\dagger&
c^{K_{T_{33}}}_{n} \left(\mathcal{R}_3\right)_n 
\end{pmatrix}
 \begin{pmatrix} T \\ T_3 \end{pmatrix} , ~~~~
\label{KTTpre}  \\
F^\dagger (\mathcal K_F - \mathds 1)  F&=& \sum_n~
 F^\dagger \left[ 
 c^{K_F}_{n}   \left(\mathcal{R}_1\right)_n
\right] F \ ,\label{KFFpre}\\
N^\dagger (\mathcal K_N - \mathds 1)  N&=& \sum_n~
 N^\dagger \left[ 
 c^{K_N}_{n}   \left(\mathcal{R}_1\right)_n
\right] N \ ,\label{KNNpre}
\end{eqnarray}%
where the $c_n$ are order one coefficients which we can assume to be real
thanks to the imposed CP symmetry. Products of flavon
fields which are allowed to couple in the K\"ahler potential are collected in
the tuples $\mathcal R_i$, which in turn are unions of tuples $\mathcal S_i$. 
These tuples, which contain all possible combinations of up to eight flavons with a
minimum contribution of order $\lambda^8$, are defined as

\be
\mathcal{R}_1=\mathcal{S}_1\cup\mathcal{S}_2\cup\mathcal{S}_3\,,\qquad
 \mathcal{R}_2=\mathcal{S}_1\cup\mathcal{S}_2\,,\qquad
 \mathcal{R}_3=\mathcal{S}_1\,,\qquad
 \mathcal{R}_4=\mathcal{S}_4\,,
\ee

\noindent where
\begin{eqnarray}
\nonumber \mathcal{S}_1&=& \bigg\{\frac{ 
   \Phi^d_2 \Phi^{d\dagger}_2}{M^2},
~\frac{\Phi^d_3\Phi^{d\dagger}_3}{M^2},
~\frac{\tilde{\Phi}^d_3\tilde{\Phi}^{d\dagger}_3}{M^2},
~\frac{\Phi^u_2\Phi^{u\dagger}_2}{M^2},
~\frac{\tilde{\Phi}^u_2\tilde{\Phi}^{u\dagger}_2}{M^2},
~\frac{(\tilde{\Phi}^u_2)^2}{M^2},
~\frac{\Phi^\nu_{3'}\Phi^{\nu\dagger}_{3'}}{M^2},
~\frac{\Phi^\nu_2\Phi^{\nu\dagger}_2}{M^2},
~\frac{\Phi^\nu_1\Phi^{\nu\dagger}_1}{M^2},
~\frac{\eta \eta^\dagger}{M^2},
\\
\nonumber &~&
~\frac{(\Phi^d_3)^2\Phi^\nu_1}{M^3},
~\frac{(\Phi^d_3)^2\Phi^\nu_2}{M^3},
~\frac{(\Phi^d_3)^2\Phi^\nu_{3'}}{M^3},
~\frac{\Phi^d_2 \Phi^{d \dagger}_2\tilde{\Phi}^u_2}{M^3},
~\frac{\Phi^{d\dagger}_2\tilde{\Phi}^{d\dagger}_3\Phi^u_2}{M^3},
~\frac{(\Phi^d_2\Phi^{d\dagger}_2)^2}{M^4},
~\frac{(\Phi^d_3\Phi^{d\dagger}_3)^2}{M^4},
\\
&~&
~\frac{\Phi^d_2 \Phi^{d \dagger}_2\Phi^d_3 \Phi^{d \dagger}_3}{M^4},
~\frac{\Phi^d_2\Phi^{d\dagger}_2\tilde{\Phi}^d_3\tilde{\Phi}^{d\dagger}_3}{M^4},
~\frac{(\Phi^d_2\Phi^{d\dagger}_2)^2\tilde{\Phi}^u_2}{M^5},
~\frac{(\Phi^d_2\Phi^{d\dagger}_2)^3}{M^6},
~\frac{(\Phi^d_2\Phi^{d\dagger}_2)^4}{M^8}
+\text{all
  h.c.}\bigg\}\,, 
\label{Operators1}\\
\mathcal{S}_2&=&\bigg\{
\frac{\tilde{\Phi}^u_2}{M},
~\frac{\Phi^\nu_{1}\Phi^{\nu\dagger}_{2}}{M^2},
~\frac{\Phi^{d\dagger}_2\tilde{\Phi}^{d\dagger}_3\Phi^u_2}{M^3}
+\text{all
  h.c.}\bigg\}\,,
\label{Operators2}\\
\mathcal{S}_3&=&\bigg\{\frac{(\Phi^d_2)^4\Phi^d_3}{M^5},
~\frac{\Phi^\nu_{1}\Phi^{\nu\dagger}_{3'}}{M^2},
~\frac{\Phi^\nu_{2}\Phi^{\nu\dagger}_{3'}}{M^2},
~\frac{\Phi^d_3\Phi^{d\dagger}_3\tilde{\Phi}^u_2}{M^3},
~\frac{(\Phi^d_2)^5\Phi^{d\dagger}_2\Phi^d_3}{M^7}
+\text{all
  h.c.}\bigg\}\,,
\label{Operators3}\\
\nonumber \mathcal{S}_4&=&\bigg\{
\frac{(\Phi^d_2)^5}{M^5},
~\frac{\eta(\Phi^{d\dagger}_2)^2}{M^3},
~\frac{\Phi^d_2\Phi^d_3\Phi^\nu_{3'}}{M^3},
~\frac{\Phi^d_2\Phi^d_3(\Phi^{d\dagger}_3)^2}{M^4},
~\frac{(\Phi^{d\dagger}_2)^2\Phi^d_3\Phi^{d\dagger}_3}{M^4},
~\frac{(\Phi^{d\dagger}_2)^3\Phi^\nu_2}{M^4},
\\
&~&
~\frac{(\Phi^{d\dagger}_2)^3(\Phi^{d\dagger}_3)^2}{M^5},
~\frac{\eta\Phi^d_2(\Phi^{d\dagger}_2)^3}{M^5},
~\frac{(\Phi^d_2)^6\Phi^{d\dagger}_2}{M^7}\bigg\}\,.
\label{Operators4}
\end{eqnarray}

\noindent $\mathcal S_1$ and $\mathcal S_2$ contain combinations of flavons
with $U(1)$ charges that sum up to zero. They can form $S_4$ invariants when
contracted with two doublets or two triplets. 
Therefore, $\mathcal{S}_1$ and $\mathcal{S}_2$ contribute to $\mathcal K_F$,
$\mathcal K_N$ and the upper-left $2\times 2$ block of $\mathcal K_T$ in
Eq.~\eqref{KTTpre}. Moreover, the combinations in $\mathcal{S}_1$ can be
contracted to $S_4$ invariants so that they additionally contribute to the lower-right
$1\times 1$ block of $\mathcal K_T$. $\mathcal S_3$ gives further
contributions to $\mathcal K_F$ and $\mathcal K_N$ but not to $\mathcal K_T$.
Finally, the combinations contained in $\mathcal S_4$  have $U(1)$ charges
which add up to $5$ and allow for $S_4$ contractions to a doublet. Hence, they
contribute to the off-diagonal upper-right block of $\mathcal K_T$. We remark
that the effects of the operators involving the flavon field~$\eta$
are independent of its $S_4$ transformation properties as a ${\bf{1}}$ or
${\bf{1'}}$. 

When calculating the K\"ahler metric from the expressions of
Eqs.~(\ref{KTTpre}-\ref{KNNpre}), it is important to take into account all
invariant $S_4$ contractions of two matter fields with a given product of
flavons.


\subsection[K\"{a}hler metric with leading order corrections]{K\"{a}hler metric with LO corrections}

It is straightforward though tedious to determine the matrices $\mathcal K_T$,
$\mathcal K_F$ and $\mathcal K_N$ from Eqs.~(\ref{KTTpre}-\ref{KNNpre}). 
Keeping only the LO corrections to the unit matrix, we find for the
${\bf{10}}$ of $SU(5)$ 
\begin{eqnarray}
\mathcal{K}_T-\mathds 1&\approx& 
\left(\begin{array}{ccc}
(k_5+k_1)\,\lambda^2&k_2\,\lambda^4&~~k_4\,e^{-i\theta^k_4}\lambda^6\\
\cdot& (k_5-k_1)\,\lambda^2&~~k_3\,e^{-i\theta^k_3}\lambda^5\\
\cdot&\cdot&~~k_6\,\lambda^2
\end{array}
\right) \ ,
\label{KTmetric}
\end{eqnarray}%
where $k_i$ denote real order one coefficients, and $\theta^k_i$ are phases
associated with the generally complex flavon VEVs. 
Here and throughout our paper, the dots in the lower-left corner of the matrix
represent the complex conjugates of the corresponding entries in the
upper-right part of the matrix. 
The operator $T^\dagger \Phi^d_2 \Phi^{d\dagger}_2T/M^2$ gives rise to the
parameters $k_1$ and $k_5$ through different $S_4$ contractions, while $k_6$
is due to $T_3^\dagger \Phi^d_2 \Phi^{d\dagger}_2T_3/M^2$. 
Being associated with $T^\dagger\tilde{\Phi}^u_2T/M$, the parameter $k_2$
carries no phase factor because $\tilde \phi^u_2\in \mathbb R$,
cf. Appendix~\ref{App:AppendixA}. Finally, the (13) and (23) elements
originate from $T^\dagger\eta(\Phi^{d\dagger}_2)^2T_3/M^3$  and 
$T^\dagger(\Phi^d_2)^5T_3/M^5$, respectively. 
Making use of the phases of the LO flavon VEVs, given explicitly in
Eq.~\eqref{eq:LOflavonphases}, we can  write the phases of
Eq.~\eqref{KTmetric} as 
\begin{eqnarray}
\theta^{k}_4=\theta^d_3-\theta^d_2
~~~~~\text{and}~~~~\theta^{k}_3&=&-5\theta^d_2 \ ,
\end{eqnarray}
where $\theta^d_2$ and $\theta^d_3$ are the phases of the LO VEVs
$\phi^d_2$ and $\phi^d_3$, respectively.

Analogously, we obtain the matrix $\mathcal K_F$,\footnote{There are also
  flavour universal $\lambda^2$ and $\lambda^4$ contributions to the diagonal
  elements of $\mathcal K_F$ which, however, do not effect our LO results.}
\begin{eqnarray}
 \mathcal{K}_F-\mathds 1 & \approx &
\left(\begin{array}{ccc}
2 K_1&K_3&K_3\\
\cdot&K_2-K_1&K_3\\
\cdot&\cdot&-(K_2+K_1)
\end{array}
\right) \lambda^4 \ ,\label{KFmetric}
\end{eqnarray}%
where $K_i\in\mathbb {R}$. The parameters on the diagonal, $K_1$ and $K_2$,
originate from different contractions of the term
$F^\dagger\Phi^d_3\Phi^{d\dagger}_3F/M^2$.  The off-diagonal elements,
parametrised by~$K_3$, 
are derived from the operator $F^\dagger \tilde{\Phi}^u_2 F/M$ and are real due to 
$\tilde \phi^u_2\in \mathbb R$.
Hence the LO correction of $\mathcal{K}_F$ from unity is given by a real matrix.

The corresponding K\"ahler metric $\mathcal K_N$ for the right-handed neutrinos
is identical to~$\mathcal K_F$ up to a difference in the order one coefficients
of the individual corrections. We thus have
\begin{eqnarray}
\mathcal{K}_N-\mathds 1&\approx&
\left(\begin{array}{ccc}
2 K^N_1&K^N_3&K^N_3\\
\cdot&K^N_2-K^N_1&K^N_3\\
\cdot&\cdot&-(K^N_2+K^N_1)
\end{array}
\right) \lambda^4 \ ,\label{KNmetric}
\end{eqnarray}%
where the coefficients $K^N_i$ are again real.


\subsection{Canonical normalisation}

The expansion of the K\"{a}hler potentials in terms of flavon insertions leads
to non-canonical kinetic terms. 
In order to bring the K\"ahler potential back to its canonical form, a
non-unitary transformation has to be applied on the matter superfields. 
This procedure is known as canonical normalisation (CN)~\cite{King:2003xq}, and
introduces the $3\times 3$ matrices $P_A$ which transform the matter
superfields $A=T,F,N$ as $A= P^{-1}_A A'$ so that 
\begin{eqnarray}
&(P_A^\dagger)^{-1}\mathcal{K}_AP_A^{-1}=\mathds 1 ~~\Longrightarrow~~
  \mathcal{K}_A=P_A^\dagger P_A \ . 
\label{Psdef} 
\end{eqnarray}%
A prescription for deriving the matrices $P_A$ can be found in
Appendix~\ref{App:Pmatrices}. To LO, they take the simple form
\begin{eqnarray}
P_T&\approx&\left(\begin{array}{ccc}
1&~~~\frac{k_2}{2}\,\lambda^4&~~~\frac{k_4}{2}e^{-i\theta^k_4}\lambda^6\\
\cdot&~~~1&~~~\frac{k_3}{2}e^{-i\theta^k_3}\lambda^5\\
\cdot&\cdot&~~~1
\end{array}
\right)  , \qquad 
P_{F(N)}\approx\left(\begin{array}{ccc}
1&~~~\frac{K^{(N)}_3}{2}\, \lambda^4&~~~\frac{K^{(N)}_3}{2}\, \lambda^4\\
\cdot&~~~1&~~~\frac{K^{(N)}_3}{2} \,\lambda^4\\
\cdot&\cdot&~~~1
\end{array}
\right)  .~~~~ \label{Ps2}
\end{eqnarray}%
In the following sections we study the structure of the Yukawa as well
as the soft supersymmetry breaking sectors. The CN transformations of
Eq.~(\ref{Ps2}) have to be applied to these before aiming at a
physical interpretation of the resulting patterns.




\section[Yukawa sector after canonical normalisation]{Yukawa sector after CN}
\label{FermionSector}
\cleqn

In this section, we study the fermionic sector of the model, completing the
analysis of~\cite {M1,M2} by including the effects of canonical normalisation. 
Our parametrisation differs slightly from the one used in~\cite {M1,M2} as, in
this work, we do not absorb any of the higher order corrections to the mass
matrices or the flavon VEVs into the associated leading order terms. See
Appendix~\ref{App:AppendixA} for more details.


\subsection{Charged fermions}
\label{ChargedFermionSector}


\subsubsection{Up-type quarks}

The Yukawa matrix of the up-type quarks can be constructed by considering all
the possible combinations of a product of flavons with $TTH_5$ for the
upper-left $2\times2$ block, with $TT_3H_5$ for the ($i3$) elements, and
with $T_3T_3H_5$ for the (33) element. The operators which generate a
contribution to the Yukawa matrix of order up to and including $\lambda^8$ are
\begin{equation}
\begin{aligned}
&y_tT_3 T_3 H_5+\frac{1}{M}y^u_1TT\Phi^u_2 H_5+\frac{1}{M^2}y^u_2TT\Phi^u_2\tilde{\Phi}^u_2 H_5\\
&+\frac{1}{M^3}y^u_{3,4}T_3 T_3(\Phi^d_3)^2\Phi^\nu_{2,3'}H_5
+\frac{1}{M^5}y^u_5TT(\Phi^d_2)^2(\Phi^d_3)^3H_5+\frac{1}{M^5}y^u_6TT_3(\Phi^d_2)^3(\Phi^d_3)^2H_5\ ,\label{YuOperators}
\end{aligned}
\end{equation}%
where the parameters $y_t$ and $y^u_{i}$ are real order one coefficients.
Inserting the flavon VEVs and expanding the $S_4$ contractions of
Eq.~\eqref{YuOperators} using the Clebsch-Gordan coefficients given for
instance in~\cite{M1}, yields the up-type Yukawa matrix at the GUT scale
\begin{eqnarray}
\mathcal Y^u_{\text{GUT}}&\approx&\left(
\begin{array}{ccc}
 y_u e^{i\theta^y_u}\lambda ^8 & 0 & 0 \\
 0 &y_c e^{i\theta^y_c} \lambda ^4&  z^u_2 e^{i\theta^{z_u}_2}\lambda ^7\\
 0 &  z^u_2 e^{i\theta^{z_u}_2}\lambda ^7 & y_t 
\end{array}
\right) \ ,\label{Yu}
\end{eqnarray}%
where the relation to the flavon VEVs,
cf. Eqs.~(\ref{LeadingFlavonVevs1}-\ref{LeadingFlavonVevs3}) as well as
Appendix~\ref{App:AppendixA}, is given by 
\be
 y_u e^{i\theta^y_u}=y^u_2\phi^u_2\tilde{\phi}^u_2
+y^u_1\delta^u_{2,1}
,\qquad
y_c \,e^{i\theta^y_c}=y^u_1\phi^u_2,
\qquad
z^u_2 e^{i\theta^{z_u}_{2}}=y^u_6(\phi^d_2)^3(\phi^d_3)^2\ .\label{dexerw1}
\ee
Applying the phases of the LO flavon VEVs as given in
Eq.~\eqref{eq:LOflavonphases}, we moreover have 
\begin{equation}
\theta^y_u=\theta^y_c=2\theta^d_2+3\theta^d_3, 
\qquad
\theta^{z_u}_2=3\theta^d_2+2\theta^d_3,\label{upphases}
\end{equation}
where we have also used the fact that the shift $\delta^u_{2,1}$ of the flavon
VEV $\vev{\Phi^u_2}$ in the first component is of order $\lambda^8$ and
proportional to $(\phi^d_2)^2 (\phi^d_3)^3$, cf. Eq.~\eqref{eq:shifts}.
It is worth noting that the (12), (13) and (21), (31) elements of Eq.~\eqref{Yu}
remain zero up to order $\lambda^8$.

Changing to the basis with canonical kinetic terms, we calculate 
$(P^{-1}_T)^T\mathcal Y^u_\text{GUT}P^{-1}_T$. For convenience we also apply an extra
phase redefinition on the
right-handed superfields,
\begin{eqnarray}
Q_u&=&\text{diag}(e^{i\theta^y_u},e^{i\theta^y_u},1).\label{Qu}
\end{eqnarray}
 As a result we obtain the up-type quark Yukawa
matrix in the canonical basis,
\begin{eqnarray}
Y^{u}_{\text{GUT}}&\approx&\left(
\begin{array}{ccc}
 y_u \,\lambda^8 & -\frac{1}{2}k_2\,y_c \,\lambda^8  &-\frac{1}{2}k_{4}\,y_te^{i\theta^k_{4}}\,\lambda ^6 \\
-\frac{1}{2}k_2\,y_c \lambda^8 & y_c\, \lambda ^4 & -\frac{1}{2}k_{3}\,y_te^{i\theta^k_{3}}\lambda^5 \\
\!\!-\frac{1}{2}k_{4}\,y_te^{i(\theta^k_{4}-\theta^y_u)}\,\lambda ^6 &~~-\frac{1}{2}k_{3}\,y_te^{i(\theta^k_{3}-\theta^y_u)}\lambda^5~~ &y_t
\end{array}
\right)\ . ~~~~
\label{YuC}
\end{eqnarray}%
Compared to Eq.~\eqref{Yu}, the canonical normalisation has significantly
modified the off-diagonal entries: the texture zeros are filled in; moreover,
the (23) and (32) elements feature a reduced $\lambda$-suppression.


\subsubsection{Down-type quarks and charged leptons}

The Yukawa matrices of the down-type quarks and the charged leptons can be
deduced from the superpotential operators
\begin{equation}
\begin{aligned}
&
y^d_1\frac{1}{M}FT_3\Phi^d_3H_{\bar{5}}+
y^d_2\frac{1}{M^2}(F\tilde{\Phi}^d_3)_1(T\Phi^d_2)_1H_{\bar{45}}
+y^d_5\frac{1}{M^3}(F(\Phi^d_2)^2)_3(T\tilde{\Phi}^d_3)_3H_{\bar{5}}\\
&
+y^d_3\frac{1}{M^2}FT_3\Phi^d_3\tilde{\Phi}^u_2H_{\bar{5}}
+y^d_4\frac{1}{M^2}FT_3\eta\tilde{\Phi}^d_3H_{\bar{5}}
+y^d_6\frac{1}{M^3}FT\Phi^d_2\tilde{\Phi}^d_3\tilde{\Phi}^u_2H_{\bar{45}}\\
&+y^d_7\frac{1}{M^5}FT(\Phi^d_2)^2(\Phi^d_3)^3H_{\bar{45}}
+y^d_8\frac{1}{M^5}FT_3(\Phi^d_2)^3(\Phi^d_3)^2H_{\bar{45}}
+y^d_9\frac{1}{M^6}FT_3(\Phi^d_2)^4(\Phi^d_3)^2H_{\bar{5}}\, ,\label{YdOperators}
\end{aligned}
\end{equation}%
where the $y^d_i$ are real order one coefficients. For the operators
proportional to $y^d_2$ and $y^d_5$, specific contractions have been chosen as
described in~\cite{M1, M2}, such that the Gatto-Sartori-Tonin
(GST)~\cite{Gatto:1968ss} and Georgi-Jarlskog (GJ)~\cite{Georgi:1979df}
relations are satisfied at LO. For all other operators we 
do not restrict the contractions to special choices; however, we have checked
that in all cases, our LO result can simply be parameterised by an effective
coupling constant which is given as a combination of the individual
contributions from each contraction. It is worth noting that the operator
proportional to $y^d_4$ is only allowed if $\eta$ transforms as a trivial
singlet under~$S_4$. Separating the contributions of $H_{\bar{5}}$  and
$H_{\bar{45}}$, the $S_4$ contractions give rise to 
\begin{equation}
\mathcal Y_{\bar{5}}\approx\left(
 \begin{array}{ccc}
 0 &\tilde{x}_2 e^{i\theta^{\tilde{x}}_2} \lambda^5&-\tilde{x}_2 e^{i\theta^{\tilde{x}}_2} \lambda^5\\
 -\tilde{x}_2 e^{i\theta^{\tilde{x}}_2} \lambda^5& 0 &\tilde{x}_2 e^{i\theta^{\tilde{x}}_2} \lambda^5 \\
 z^d_3 e^{i\theta^{z_d}_3} \lambda^6&z^d_2 e^{i\theta^{z_d}_2} \lambda^6 &y_be^{i\theta^y_b} \lambda^2
\end{array}
\right)  ,~\quad
\mathcal Y_{\bar{45}} \approx \left(
\begin{array}{ccc}
 z^d_1e^{i\theta^{z_d}_1}\lambda^8&0&0\\
0& y_s e^{i\theta^y_s}\lambda^4 &-y_s e^{i\theta^y_s}\lambda^4 \\
 0 & 0&0
\end{array}
\right)  .~ \label{Ydb}
\end{equation}
The parameters in these expressions are related to the flavon VEVs
as defined in Eqs.~(\ref{LeadingFlavonVevs1}-\ref{LeadingFlavonVevs3}) and
Appendix~\ref{App:AppendixA} via
\be
\nonumber
\tilde{x}_2e^{i\theta^{\tilde{x}}_2}=y^d_5(\phi^d_2)^2\tilde{\phi}^d_3\, , ~\quad
y_be^{i\theta^y_b}=y^d_1\phi^d_3\, ,~\quad
z^d_2e^{i\theta^{z_d}_2}=y^d_1\delta^d_{3,3}\!+y^d_3\phi^d_3\tilde{\phi}^u_2\, ,~\quad
z^d_3e^{i\theta^{z_d}_3}=y^d_1\delta^d_{3,1}\, , 
\ee
\be
y_se^{i\theta^y_s}=y^d_2\phi^d_2\tilde{\phi}^d_3\ ,~\quad
z^d_1e^{i\theta^{z_d}_1}=
y^d_{7}(\phi^d_2)^2(\phi^d_3)^3-y^d_{6}\phi^d_2\tilde{\phi}^d_3\tilde{\phi}^u_2
\ .\label{ax}
\ee
Using Eqs.~(\ref{eq:LOflavonphases},\ref{eq:phasesofshifts}), we deduce the following relations for the phases
\be
\theta^{\tilde{x}}_2=3(\theta^d_2+\theta^d_3)\ ,~\quad
\theta^y_s=\theta^{z_d}_1=2\theta^d_2+3\theta^d_3\ ,~\quad 
\theta^y_b=\theta^{z_d}_2=\theta^{z_d}_3=\theta^d_3\ .\label{downphases}
\ee
The Yukawa matrices of the down-type quarks and the charged leptons 
are linear combinations of the two structures in
Eq.~(\ref{Ydb}). Following the construction proposed by Georgi
and Jarlskog, we have
$\mathcal Y^d_{\text{GUT}}=\mathcal Y_{\bar{5}}+\mathcal Y_{\bar{45}}$ and
$\mathcal Y^e_{\text{GUT}}=(\mathcal Y_{\bar{5}}-3\mathcal Y_{\bar{45}})^T$,
respectively. 

Performing the canonical normalisation on the Yukawa matrices
$(P^{-1}_T)^T\mathcal Y^d_{{\text{GUT}}}P^{-1}_F$ and
$(P^{-1}_F)^T \mathcal Y^{e}_{{\text{GUT}}}P^{-1}_T$ as well as an additional
rephasing of the right-handed superfields by
\begin{eqnarray}
Q_{d}&=&Q_{e}=\text{diag}(e^{i\theta^{\tilde x}_2},e^{i\theta^{\tilde
    x}_2},e^{i\theta^y_b}),\label{Qd}
\end{eqnarray}
 we end up with\\[-9mm]

{\small{
\begin{eqnarray}
 Y^{d}_{{\text{GUT}}}& \!\approx\!&\left(
\begin{array}{ccc}
 e^{i(\theta^{z_d}_1-\theta^{\tilde{x}}_2)}z^d_1\lambda^8&\tilde x_2\lambda^5&\!-e^{i(\theta^{\tilde{x}}_2-\theta^y_b)}\tilde{x}_2 \lambda^5\\
 -\tilde{x}_2\lambda^5&~~e^{i(\theta^y_s-\theta^{\tilde{x}}_2)}y_s\lambda^4&\!-e^{i(\theta^y_s-\theta^y_b)}y_s\lambda^4\\
\!e^{-i\theta^{\tilde{x}}_2}\left(z^d_{3}e^{i\theta^{z_d}_3}\!-\!\frac{K_{3}}{2}e^{i\theta^y_b}y_b\right)\lambda^6\!&~e^{-i\theta^{\tilde{x}}_2}\left(z^d_{2}e^{i\theta^{z_d}_2}\!-\!\frac{K_{3}}{2}e^{i\theta^y_b}y_b\right)\lambda^6\!\!&y_b \lambda^2
\end{array}
\right) \!, ~~~~~~~~\label{YdC}\\
 Y^{e}_{{\text{GUT}}}&\!\approx\!&\left(
\begin{array}{ccc}
 -3e^{i(\theta^{z_d}_1-\theta^{\tilde{x}}_2)}y_d\lambda^8&-\tilde{x}_2\lambda^5&~~e^{-i\theta^y_b}\left(z^d_{3}e^{i\theta^{z_d}_3}-\frac{K_{3}}{2}e^{i\theta^y_b}y_b\right)\lambda^6 \\
 \tilde{x}_2\lambda^5&-3\,e^{i(\theta^y_s-\theta^{\tilde{x}}_2)}y_s\lambda^4&~~e^{-i\theta^y_b}\left(z^d_{2}e^{i\theta^{z_d}_2}-\frac{K_{3}}{2}e^{i\theta^y_b}y_b\right)\lambda^6\\
-\tilde{x}_2\lambda^5&3\,e^{i(\theta^y_s-\theta^{\tilde{x}}_2)}y_s\lambda^4&~~y_b \lambda^2
\end{array}
\right) \!.\label{YeC}
\end{eqnarray}}}%

\noindent We observe that the canonical normalisation modifies the down-type quark and
charged lepton Yukawa matrices solely by additional contributions of the same
order in the (31), (32) and (13), (23) elements, respectively. 
Comparing Eq,~\eqref{YdC} with Eq.~\eqref{YuC} suggests that the CKM mixing is
dominated by the diagonalisation of the down-type quark Yukawa matrix. We will
explicitly verify this when calculating the SCKM transformations in
Section~\ref{section:SCKM}.


\subsection{Neutrinos}


\subsubsection{Dirac neutrino coupling}

Having introduced  right-handed neutrinos $N$ in Table~\ref{particles}, their
Dirac coupling to the left-handed SM neutrinos originates from the
superpotential terms
\begin{equation}
\begin{aligned}
&y_D FNH_5+y^D_1\frac{1}{M}FN\tilde{\Phi}^u_2H_5+y^D_2\frac{1}{M^2}FN(\tilde{\Phi}^u_2)^2H_5+y^D_{3,4,5}\frac{1}{M^3}FN(\Phi^d_3)^2\Phi^{\nu}_{1,2,3'}H_5\\
&+y^D_6\frac{1}{M^5}FN(\Phi^d_2)^4\Phi^d_3H_5,\label{YnuOperators}
\end{aligned}
\end{equation}%
where $y_D$ and $y^D_i$ are real order one parameters. The corresponding Yukawa
matrix is determined as
\begin{eqnarray}
 \mathcal Y^\nu%
&\approx&\left(
\begin{array}{ccc}
y_D&~~~z^D_{2}e^{i\theta^{z_D}_2}\lambda^6&z^D_{1}\,\lambda^4\\
z^D_{2}e^{i\theta^{z_D}_2}\lambda^6&z^D_{1}\,\lambda^4&y_D\\
z^D_{1}\,\lambda^4&y_D&~~~z^D_{2}e^{i\theta^{z_D}_2}\lambda^6
\end{array}
\right) \ , \label{Ynu}
\end{eqnarray}%
with 
\be
 z^D_{1}=y^D_1\tilde{\phi}^u_2\ ,\qquad
z^D_{2}e^{i\theta^{z_D}_{2}}=y^D_1\tilde{\delta}^u_{2,1} \ ,\qquad
\theta^{z_D}_{2}=4\theta^d_2+\theta^d_3 \ . \label{zD1}
\ee
Here, the phase can be deduced from Eq.~\eqref{eq:shifts}.

Applying the CN transformation $(P^{-1}_F)^T\mathcal Y^\nu
P^{-1}_N$, the corresponding Yukawa
matrix in the basis with canonical kinetic terms takes the form
\begin{eqnarray}\label{YnuC}
Y^{\nu}%
&\!\approx\!&\left(
\begin{array}{ccc}
y_D&-\frac{y_D(K_{3}+K^N_3)}{2}\lambda^4&\left(z^D_1-\frac{y_D(K_{3}+K^N_3)}{2}\right)\lambda^4\\
-\frac{y_D(K_{3}+K^N_3)}{2}\lambda^4&\left(z^D_1-\frac{y_D(K_{3}+K^N_3)}{2}\right)\lambda^4\!\!&y_D\\
\!\left(z^D_1-\frac{y_D(K_{3}+K^N_3)}{2}\right)\lambda^4&y_D&-\frac{y_D(K_{3}+K^N_3)}{2}\lambda^4
\end{array}
\right) . ~~~~~~~~~ 
\end{eqnarray}%
Compared to Eq.~\eqref{Ynu}, an additional contribution of the same order
arises in the (13), (22) and (31) entries. Moreover, the $\lambda$-suppression of
the (12), (21) and (33) elements is reduced.


\subsubsection{Majorana neutrino mass}

The mass matrix of the right-handed neutrinos is obtained from the
superpotential terms
\begin{equation}
w_{1,2,3}NN\Phi^\nu_{1,2,3'}+w_4\frac{1}{M}NN\Phi^d_2\eta+w_{5,6,7}\frac{1}{M}NN\tilde{\Phi}^u_2\Phi^\nu_{1,2,3'}+w_8\frac{1}{M^7}NN(\Phi^d_2)^8\ , 
\label{YNOperators}
\end{equation}%
where  $w_i$ denote real order one coefficients.  This results in a
right-handed Majorana neutrino mass matrix $\mathcal M_R$ of the form
\begin{eqnarray}
\frac{\mathcal M_R}{M}&\approx&\left(
\begin{array}{ccc}
A+2C&~~~B-C &~~~B-C\\
B-C&~~~B+2C &~~~A-C\\
B-C&~~~A-C&~~~B+2C
\end{array}
\right)e^{i\theta_A}\lambda^4+
\left(\begin{array}{ccc}
0&~0&~D\\
0&~D&~0\\
D&~0&~0
\end{array}
\right)e^{i\theta_D}\lambda^5 \, , ~~~~~~~ \label{MR}
\end{eqnarray}%
with 
\be
Ae^{i\theta_{A}}\!=w_1\phi^\nu_1\, , \,\quad
Be^{i\theta_A}\!=w_2\phi^\nu_2\, , \,\quad
Ce^{i\theta_A}\!=w_3\phi^\nu_{3'}\, , \,\quad
De^{i\theta_D}\!=w_2(\delta^\nu_{2,1}\!\!-\delta^\nu_{2,2})+
w_4\,\eta\,\phi^d_2\,. \label{ABCD}
\ee
According to
Eqs.~(\ref{eq:LOflavonphases},\ref{eq:shifts},\ref{eq:phasesofshifts}), the
phases are given by
\be
\theta_A=-2\theta^d_3\ ,\qquad
\theta_D=4\theta^d_2-\theta^d_3\ .
\ee
The first matrix of Eq.~\eqref{MR} arises from terms involving only
$\Phi^\nu_{1,2,3'}$. As their VEVs respect the tri-bimaximal (TB) Klein symmetry 
 $Z_2^S\times Z_2^U\subset S_4$, this part is of TB form. The second matrix of
Eq.~\eqref{MR}, proportional to $D$, is due to the operator
$w_4\frac{1}{M}NN\Phi^d_2\eta$. As the product of both flavon VEVs involved is
not an eigenvector of $U$, half of the TB Klein symmetry is broken at a relative
order of $\lambda$. The resulting trimaximal TM$_2$~\cite{TM2mix} structure can
accommodate the sizable value of the reactor neutrino mixing angle
$\theta^l_{13}$ as explained in~\cite{M2} in the context of the original
model~\cite{M1}. 

Performing the CN basis transformation $(P^{-1}_N)^T\mathcal M_R P^{-1}_N$
does not alter the matrix in Eq.~\eqref{MR} at the given order, so that 
$M_R = \mathcal M_R   + \mathcal O(\lambda^6) M$.


\subsubsection{Effective light neutrino mass matrix}

Calculating the effective light neutrino mass matrix which arises via the
type~I seesaw mechanism $v_u^2\, Y^\nu M_R^{-1} (Y^\nu)^T$, 
we can parameterise the LO result as
\begin{eqnarray}\label{eq:efflightnu}
 m^{\text{eff}}_{\nu}&\approx&\frac{y_D^2\upsilon_u^2}{\lambda^4 M}\left[ \! \left(
\begin{array}{ccc}
b^\nu +c^\nu-a^\nu &a^\nu&a^\nu\\
a^\nu&b^\nu &c^\nu \\
a^\nu &c^\nu&b^\nu
\end{array}
\right)e^{-i\theta_A}+\left(
\begin{array}{ccc}
0&0&d^\nu\\
0&d^\nu&0\\
d^\nu &0&0
\end{array}
\right)\lambda\,e^{i(\theta_D-2\theta_A)} \right]  ,~~~~~~
\end{eqnarray}%
with $a^\nu$, $b^\nu$, $c^\nu$ and $d^\nu$ being functions of the real parameters
$A$, $B$, $C$ and $D$. The deviation from tri-bimaximal neutrino mixing is controlled
by $d^\nu \propto D$. Due to the three independent LO input parameters
($w_1\propto A$ ,$w_2\propto B$ ,$w_3\propto C$), any neutrino mass spectrum
can be accommodated in this model. At this order, the canonical normalisation
does not modify the effective light neutrino mass matrix as obtained without
the CN transformations. Hence, concerning the results on light neutrino masses
and mixing, we can simply refer the reader to the corresponding discussion
in~\cite{M2}.




\section[Soft SUSY breaking sector after canonical normalisation]{Soft SUSY breaking sector after CN}
\label{sec:soft}
\cleqn

Having applied the CN basis transformation of the matter superfields to the
Yukawa sector, we now turn to the soft SUSY breaking terms. In the context of
the general MSSM with $R$-parity, these are parameterised as~\cite{Chung:2003fi}
\begin{eqnarray}\label{eq:softL}
\nonumber -\mathcal{L}_\text{soft}&\supset&  H_u \tilde{Q}_i
 A^u_{ij}\tilde{u}^c_j+H_d \tilde{Q}_i  A^d_{ij}\tilde{d}^c_j+H_d \tilde{L}_i
 A^e_{ij}\tilde{e}^c_j +H_u \tilde{L}_i  A^\nu_{ij}\tilde{N}_j + \text{h.c.} \\
\nonumber &+&\tilde{Q}^\alpha_i m^2_{Q_{ij}} \tilde{Q}^{\alpha
  *}_j+\tilde{L}^\alpha_i m^2_{L_{ij}} \tilde{L}^{\alpha *}_j+\tilde{u}^{c*}_i
m^2_{u^c_{ij}} \tilde{u}^c_j+\tilde{d}^{c*}_i m^2_{d^c_{ij}}
\tilde{d}^c_j+\tilde{e}^{c*}_i m^2_{e^c_{ij}} \tilde{e}^c_j+\tilde{N}^{*}_i
m^2_{N_{ij}} \tilde{N}_j\\
&+&m^2_{H_u}|H_u|^2+m^2_{H_d}|H_d|^2\ ,
\end{eqnarray}%
and contain trilinear scalar couplings ($A$-terms) as well as bilinear scalar
masses. A tilde indicates the scalar partner $\tilde f$ of a SM fermion $f$.
Taking into account the $SU(5)$ framework, we construct the effective soft
SUSY breaking operators in this section, assuming that the mechanism of SUSY
breaking is practically independent of the family symmetry breaking.


\subsection{Trilinear soft couplings}

The flavour structure of the trilinear $A$-terms is similar to the
corresponding Yukawa matrices, as both originate from the same set of
superpotential terms. In the case of the soft terms, these are coupled to a
hidden sector superfield $X$ with independent real order one coupling
constants and suppressed by a mass scale $M_X$. When $X$ develops its SUSY
breaking $F$-term VEV, the scalar components of the Higgs and matter
superfields are projected out, thereby generating the trilinear soft terms. 
There exist in fact extra contributions to the $A$-terms from superpotential
operators involving flavons but no $X$~field. These can be traced back to
non-vanishing VEVs for the auxiliary $F$-components of the flavon fields, which
are zero in the SUSY limit but develop a non-trivial value when SUSY breaking
terms are included. It turns out that such $F$-term VEVs are aligned with the LO
flavon VEVs in many situations~\cite{Ross:2002mr,Feruglio:2009iu}. Hence,
these extra contributions to the $A$-terms do not give rise to new flavour structures.

Defining the mass parameters $m_0\equiv \vev{F_X} /M_X$ and $A_0\equiv
\alpha_0\, m_0$, with $\alpha_0$ being a real constant,  we can obtain the
expressions for the trilinear matrices $\mathcal  A^f_{\text{GUT}}/A_0$ by
copying the Yukawas matrices of Eqs.~(\ref{Yu},\ref{Ydb},\ref{Ynu}) with 
different order-one coefficients and phases: 
$y_f\to a_f$, 
$\tilde{x}_2\to \tilde{x}^a_2$, 
$z^f_i\to z^{f_a}_i$, 
$y_D\to\alpha_D$ as well as
$\theta^y_f\to\theta^a_f$, 
$\theta^{\tilde{x}}_2\to\theta^{\tilde{x}_a}_2$, 
$\theta^{z_{f}}_i\to \theta^{z_{f_a}}_i$. 
With these replacements, we find
\begin{eqnarray}
\frac{\mathcal A^u_{\text{GUT}}}{A_0}&\approx&\left(
\begin{array}{ccc}
 a_u \, e^{i\theta^a_u}\lambda ^8 & 0 & 0 \\
 0 &a_c \, e^{i\theta^a_c} \lambda ^4& z^{u_a}_2 e^{i\theta^{z_{u_a}}_2}\lambda ^7\\
 0 & z^{u_a}_2 e^{i\theta^{z_{u_a}}_2}\lambda ^7 & a_t 
\end{array}
\right) \ , \label{Au}
\end{eqnarray}%
and similarly for $\mathcal A^d_{\text{GUT}}$, $\mathcal A^e_{\text{GUT}}$ and
$\mathcal A^\nu$.
Applying the CN transformation as well as the rephasing of the right-handed
superfields proceeds analogously to the Yukawa sector. The resulting trilinear
matrices $A^f_{\text{GUT}}/A_0$ in the basis of canonical kinetic terms are thus
derived from Eqs.~(\ref{YuC},\ref{YdC},\ref{YeC},\ref{YnuC}) by simply replacing
$y_u\to a_u\, e^{i(\theta^a_u-\theta^y_u)}$, 
$y_c\to a_c\,e^{i(\theta^a_c-\theta^y_u)}$, 
$y_t\to a_t$, 
$y_s\to a_s\,e^{i(\theta^a_s-\theta^y_s)}$, 
$y_b\to a_b\,e^{i(\theta^a_b-\theta^y_b)}$, 
$\tilde{x}_2\to
\tilde{x}^a_2\,e^{i(\theta^{\tilde{x}_a}_2-\theta^{\tilde{x}}_2)}$, 
$z^f_i\to z^{f_a}_i\,e^{i(\theta^{z_{f_a}}_i-\theta^{z_{f}}_i)}$ and 
$y_D\to\alpha_D$.
For example, the up-type quark trilinear matrix takes the form
\begin{eqnarray}
\frac{A^{u}_{\text{GUT}}}{A_0}&\approx&\left(
\begin{array}{ccc}
a_u\, e^{i(\theta^a_u-\theta^y_u)}  \,\lambda^8 & 
-\frac{1}{2}k_2\,a_c\,e^{i(\theta^a_c-\theta^y_u)} \,\lambda^8  &
-\frac{1}{2}k_{4}\,a_te^{i\theta^k_{4}}\,\lambda ^6 \\
-\frac{1}{2}k_2\,a_c\,e^{i(\theta^a_c-\theta^y_u)} \lambda^8 & 
a_c\,e^{i(\theta^a_c-\theta^y_u)}\, \lambda ^4 & 
-\frac{1}{2}k_{3}\,a_te^{i\theta^k_{3}}\lambda^5 \\
\!\!-\frac{1}{2}k_{4}\,a_te^{i(\theta^k_{4}-\theta^y_u)}\,\lambda ^6 &~~-\frac{1}{2}k_{3}\,a_te^{i(\theta^k_{3}-\theta^y_u)}\lambda^5~~ &a_t
\end{array}
\right) \ . ~~~~
\label{AuC}
\end{eqnarray}%


\subsection{Soft scalar masses}

The scalar mass terms of the soft supersymmetry breaking Lagrangian originate
from the K\"ahler potential. Non-renormalisable couplings of the matter
superfields to the square $X^\dagger X/M_X^2$ of the SUSY breaking field $X$
generate soft masses when the $F$-term of $X$ develops a VEV. 
The structure of the soft mass matrices is therefore similar to the K\"ahler
metric $\mathcal K$ of the corresponding GUT multiplet. As for the trilinear
soft terms, all order one coefficients are independent of those appearing
in~$\mathcal K$. The scalar masses before canonical normalisation are then obtained
from $\mathcal K_T$, $\mathcal K_F$ and $\mathcal K_N$ of
Eqs.~(\ref{KTmetric},\ref{KFmetric},\ref{KNmetric}) by
replacing
$k_i\to b_i$, 
$\theta^k_i\to \theta^b_i$,
$K_i\to B_i$ and
$K^N_i\to B^N_i$.
Moreover, the ones on the diagonal of $\mathcal K$ have to be rescaled by a
new factor of order one. In the case of the ${\bf{10}}$ of $SU(5)$, the 2+1
structure requires the introduction of two extra parameters,
$b_{01}$ and $b_{02}$. Explicitly, we get
\begin{eqnarray}
\frac{\mathcal M^2_{T_{\text{GUT}}}}{m_0^2}&\!\approx\!&\left(\begin{array}{ccc}
b_{01}+(b_5+b_1)\lambda^2&~~~ b_{2}
\lambda^4
&~~~b_{4}\,e^{-i\theta^k_4}\lambda ^6\\
\cdot &~~~b_{01}+(b_5-b_1)\lambda^2&~~~b_{3}\,e^{-i\theta^k_3}\lambda ^5\\
\cdot &\cdot &~~~b_{02}+b_6\lambda^2
\end{array}
\right)\ ,
\label{MT} 
\end{eqnarray}%
\begin{eqnarray}
 \frac{\mathcal M^2_{F(N)_{\text{GUT}}}}{m_0^2}&\!\approx\!&\left(\begin{array}{ccc}
B^{(N)}_0\!+2B^{(N)}_1 \lambda^4 &B^{(N)}_{3}
\lambda^4~&~B^{(N)}_{3}
\lambda^4\\
\cdot&\!\!\!B^{(N)}_0 \!+(B^{(N)}_2\!-B^{(N)}_1)\lambda^4\!\!\! &~B^{(N)}_{3}
\lambda^4\\
\cdot&\cdot&\!\!\!B^{(N)}_0\!-(B^{(N)}_2\!+B^{(N)}_1)\lambda^4
\end{array}
\right) \ .~~~~~~ \label{MF}
\end{eqnarray}%
Performing the transformations to the basis of canonical kinetic terms results
in soft scalar mass matrices of the form
\begin{eqnarray}
\frac{M^2_{T_{\text{GUT}}}}{m_0^2}&\approx&
\left(
\begin{array}{ccc}
 b_{01} & ~~(b_2-b_{01}k_2)\lambda^4 & ~~~e^{-i\theta^k_4}(b_4-\frac{k_4(b_{01}+b_{02})}{2})\lambda^6 \\
\cdot & ~~b_{01} &~~~e^{-i\theta^k_3}(b_3-\frac{k_3(b_{01}+b_{02})}{2})\lambda^5 \\
\cdot & \cdot &~~ b_{02}
\end{array}
\right) \ , \label{MTC}\\
 \frac{M^2_{F(N)_{\text{GUT}}}}{m_0^2}&\approx&
\left(
\begin{array}{ccc}
 B^{(N)}_0 & ~~~(B^{(N)}_{3}
-K^{(N)}_3)\lambda^4&~~~(B^{(N)}_{3}
-K^{(N)}_3)\lambda^4 \\
 \cdot& ~~~B^{(N)}_0 &~~~ (B^{(N)}_{3}
-K^{(N)}_3)\lambda^4\\
 \cdot & \cdot & B^{(N)}_0
\end{array}
\right) \ . \label{MFC}
\end{eqnarray}%
For convenience, we will absorb the order one parameter $B_0$ into the
soft SUSY breaking mass $m_0$, so that the leading contribution on the
diagonal of $M^2_{F_{\text{GUT}}}/m_0^2$ is nothing but unity.
For the right-handed fields contained in the GUT multiplets, an additional
rephasing has to be applied. We will come back to this when calculating the
soft terms in the SCKM basis in Section~\ref{softSCKM}.
Notice that we have dropped all $\lambda$-suppressed corrections of the diagonal
elements. This simplification is justified as FCNC processes are induced by
loop diagrams involving the off-diagonal entries of the sfermion mass
matrices. The simplification of the diagonal elements in
Eqs.~(\ref{MTC},\ref{MFC}) does not affect these off-diagonals in our LO
analysis, even when going to the SCKM basis.




\section{SCKM basis}\label{section:SCKM}
\cleqn

Predictions relating a theoretical model with its phenomenological implications
are typically given in the basis in which the Yukawa matrices are diagonal
and positive, corresponding to the physical quark and lepton mass eigenstates.
The so-called SCKM basis is the analogue in a supersymmetric
framework. Changing to the SCKM basis, all canonically normalised quantities
undergo a unitary transformation of the superfields which diagonalises the
effective Yukawa couplings in the superpotential. In this basis it is
convenient to define a set of dimensionless parameters, known as the ``mass
insertion parameters'', which directly enter the expressions of
phenomenological flavour observables.

 In principle, the SCKM transformation should be performed after
 electroweak symmetry breaking. The canonically normalised Yukawa, trilinear
 and soft mass matrices should be evolved from the GUT scale $M_\text{GUT}$ to the
 weak scale $M_W$ using the corresponding renormalisation group equations
 (RGEs). Only at that point, the diagonalisation of the Yukawa matrices should
 take place, leading to the definition of a SCKM basis. Following this
 procedure, there is obviously no notion of mass insertion parameters at the scale
 $M_\text{GUT}$ as there is no proper definition of the SCKM basis.

An alternative approach which is commonly used consists in diagonalising the
Yukawa matrices at (or rather just below) the GUT scale. 
The so-obtained basis is approximately identical to the SCKM basis provided
the RGE contributions to the off-diagonal elements of the Yukawa matrices
remain negligible.\footnote{For the charged fermion sector, this
is a valid approximation thanks to the hierarchical masses of quarks and
charged leptons. In the neutrino sector, RGE contributions can be sizable in
supersymmetric models with large $t_\beta$ and a quasi-degenerate neutrino
mass spectrum~\cite{Antusch:2003kp}. They are however negligible for small
$t_\beta$ [which is realised in our scenario due to the suppression of the bottom
Yukawa coupling by two powers of $\lambda$, see Eq.~\eqref{YdC}] and a normal
neutrino mass hierarchy [which we assume in the following].} 
This is the case as long as the RGE effects can be absorbed
into a redefinition of the (unknown) order one coefficients. It is then
possible to introduce mass insertion parameters already at $M_\text{GUT}$. Their
low energy values have to be determined from the corresponding RG evolution.
In this work, we will adopt the latter approach as it allows for a semi-analytical
study of the relations between the high and low energy parameters by means of
a perturbative $\lambda$-expansion.


\subsection{SCKM transformations}

The SCKM transformations are applied on the matter superfields 
${\hat f_{L,R} \to   U_{L,R}^f \hat f_{L,R}}$,
where $U^f_{L,R}$ denote unitary $3\times 3$ matrices. These diagonalise the
canonically normalised Yukawa matrices $Y^f$ 
\begin{equation}
 (U^f_L)^\dagger Y^f U^f_R~=~\tilde{Y}^f_\text{diag} \ ,
\end{equation}%
where we use the tilde to denote the SCKM basis.
The derivation and the explicit form of the unitary transformations can be
found in Appendix~\ref{App:SCMKMatrices}. Applying this change of basis to
the Yukawa matrices yields
\begin{equation} 
\tilde{Y}^u_{\text{GUT}}\approx
\left(
\begin{array}{ccc}
y_u\lambda ^8 & 0 & 0 \\
 0 &y_c \lambda ^4& 0 \\
 0 & 0 &y_t
\end{array}
\right)  , \qquad
\tilde{Y}^d_{\text{GUT}}\approx
\left(
\begin{array}{ccc}
\frac{\tilde{x}_2^2}{y_s}\lambda^6& 0 & 0 \\
 0 &y_s\lambda ^4& 0 \\
 0 & 0 &y_b\lambda ^2
\end{array}
\right) , \label{Yudt}
\end{equation}

\begin{equation}
\tilde{Y}^e_{\text{GUT}}\approx
\left(
\begin{array}{ccc}
\frac{\tilde{x}_2^2}{3y_s}\lambda^6& 0 & 0 \\
 0 &3y_s\lambda ^4& 0 \\
 0 & 0 &y_b\lambda^2
\end{array}
\right)  . \label{Yet}
\end{equation}%
These results, which are valid at the high scale, agree with the LO results
derived in~\cite{M1,M2}. This shows that the canonical normalisation does not
affect the LO expressions of the quark and charged lepton masses. 

Up to phase convention, the CKM matrix is given by 
$V_{\text{CKM}_{\text{GUT}}}=(U^u_L)^TU^{d*}_L$ 
(see Appendix~\ref{App:SCMKMatrices} for explicit expressions). Extracting the
mixing angles 
\begin{equation}
 \sin(\theta^q_{13})_{\text{GUT}}\approx
\frac{\tilde{x}_2}{y_b}\lambda^3\ ,\qquad
\tan(\theta^q_{23})_{\text{GUT}}\approx
\frac{y_s}{y_b}\lambda^2\ ,\qquad
\tan(\theta^q_{12})_{\text{GUT}}\approx
\frac{\tilde{x}_2}{y_s}\lambda \ ,
\end{equation}%
shows that the LO CKM mixing arises purely from the down-type quark sector,
incorporating the GST relation~\cite{Gatto:1968ss} 
$\theta^q_{12}
\approx\sqrt{m_d/m_s}$, 
and agrees with the results obtained in~\cite{M1,M2}.
Concerning the CP violation, we find the Jarlskog
invariant~\cite{Jarlskog:1985ht} to be 
\begin{eqnarray}
 J^q_{\text{CP}_{\text{GUT}}}
&\approx&\lambda^7\frac{\tilde{x}_2^3}{y_b^2y_s}\sin(\theta^d_2)
 \ .
\label{JCPq}
\end{eqnarray}%

The PMNS matrix is dominated by the trimaximal TM$_2$ neutrino mixing $V_\nu$
which diagonalises the effective light neutrino mass matrix of
Eq.~\eqref{eq:efflightnu}.  Including the charged lepton corrections, we have
$U_{\text{PMNS}_{\text{GUT}}}=(U^e_L)^TV_\nu^*$ with mixing angles given as
\begin{eqnarray}
\tan(\theta^l_{23})_{\text{GUT}}&\approx&
1+\lambda\,\frac{d^\nu}{2(a^\nu-c^\nu)}\cos(4\theta^d_2+\theta^d_3) \ ,\\
\tan(\theta^l_{12})_{\text{GUT}}&\approx&
\frac{1}{\sqrt{2}}-\lambda\frac{\tilde x_2}{2\sqrt{2}y_s}\cos(\theta^d_2)\ ,\\
\sin(\theta^l_{13})_{\text{GUT}}&\approx&
\frac{\lambda}{6\sqrt{2}y_s}
\Bigg[\left(\frac{3d^\nu
  y_s\cos(4\theta^d_2+\theta^d_3)+2(a^\nu-c^\nu)\tilde x_2\cos(\theta^d_2)}{a^\nu-c^\nu}\right)^2
\nonumber
\\
&&~~~~~~~+\left(\frac{3d^\nu
  y_s\sin(4\theta^d_2+\theta^d_3)+2(a^\nu-b^\nu)\tilde x_2\sin(\theta^d_2)}{a^\nu-b^\nu}\right)^2\Bigg]^{\frac{1}{2}}
\! \ ,~~~~~~~
\end{eqnarray}%
and a leptonic Jarlskog invariant of the form
\begin{eqnarray}
\nonumber J^l_{\text{CP}_{\text{GUT}}}&\approx&
-\frac{\lambda}{36}\left(\frac{2\tilde x_2}{y_s}\sin(\theta^d_2)+\frac{3d^\nu}{a^\nu-
  b^\nu}\sin(4\theta^d_2+\theta^d_3)\right) \ .~~~\label{JCPl}
\end{eqnarray}%


\subsection{Soft terms in the SCKM basis}\label{softSCKM}

In order to obtain the flavour structure of the soft SUSY breaking terms in
a basis which is suitable for physical interpretations, we have to apply the
SCKM transformations on the canonical trilinear soft couplings and soft scalar
masses, cf.~Section~\ref{sec:soft}. The action of the $U^f_{L,R}$ matrices
on the $A$-terms is identical to the transformation of the Yukawa
matrices:

\vspace{-1.5mm}

\begin{equation}
 (U^f_L)^\dagger A^f_{\text{GUT}} U^f_R~=~\tilde{A}^f_\text{GUT}.
\end{equation}\\[-3.5mm]
 However, due to different order one coefficients, the $A$-terms
remain non-diagonal in the SCKM basis. 
The soft masses of Eqs.~(\ref{MTC},\ref{MFC}) are transformed differently for
different components of the $SU(5)$ multiplets. Moreover, we have to
associate the mass matrices of the effective soft Lagrangian in
Eq.~\eqref{eq:softL} with $M^2_{T_{\text{GUT}}}$ and $M^2_{F_{\text{GUT}}}$
and take into account the additional rephasing transformations of the
right-handed superfields, see Eqs.~(\ref{Qu},\ref{Qd}), that were performed
after CN. Then, the soft masses in the SCKM basis are
\begin{eqnarray}
(\tilde{m}_{u}^2)_{{LL}_{\text{GUT}}}&=&(U^u_L)^\dagger M^{2~*}_{T_{\text{GUT}}} U^u_L,~~~~
(\tilde{m}_{u}^2)_{{RR}_{\text{GUT}}}=(U^u_R)^\dagger Q_u\,M^2_{T_{\text{GUT}}} Q_u^\dagger\, U^u_R,\\
(\tilde{m}_{d}^2)_{{LL}_{\text{GUT}}}&=&(U^d_L)^\dagger M^{2~*}_{T_{\text{GUT}}} U^d_L,~~~~
(\tilde{m}_{d}^2)_{{RR}_{\text{GUT}}}=(U^d_R)^\dagger Q_{d}\,M^2_{F_{\text{GUT}}} Q_{d}^\dagger \,U^d_R,\\
(\tilde{m}_{e}^2)_{{LL}_{\text{GUT}}}&=&(U^e_L)^\dagger M^{2~*}_{F_{\text{GUT}}} U^e_L,~~~~
(\tilde{m}_{e}^2)_{{RR}_{\text{GUT}}}=(U^e_R)^\dagger Q_{d}\,M^2_{T_{\text{GUT}}} Q_{d}^\dagger \,U^e_R.
\end{eqnarray}

We find the following leading order expressions, where the
order one coefficients are defined in Eqs.~(\ref{Bt},\ref{adt}). Note that
we have absorbed the order one coefficient $B_0$ into~$m_0$,
cf. Eq.~\eqref{MFC}, so that ${(\tilde{m}_d^2)_{{RR}_{\text{GUT}}}}/m_0^2$ and
$(\tilde{m}_e^2)_{{LL}_{\text{GUT}}}/m_0^2$ have 1s on the diagonal.


\subsubsection*{Up-type quark sector:}

\begin{eqnarray}
\frac{\tilde{A}^u_\text{GUT}}{A_0}&\approx&\left(
\begin{array}{ccc}
\tilde{a}^u_{11}\lambda^8  & 0 & 0 \\
0&\tilde{a}^u_{22}\lambda ^4&e^{i\theta^d_2}\tilde{a}^u_{23} \lambda ^7 \\
0&e^{i(3\theta^d_2+\theta^d_3)}\tilde{a}^u_{23} \lambda ^7&\tilde{a}^u_{33}
\end{array}
\right)\ ,\label{AutGUT}\\
\frac{(\tilde{m}_{u}^2)_{{LL}_{\text{GUT}}}}{m_0^2}&\approx&\left(
\begin{array}{ccc}
 b_{01} &~~~~e^{-i\theta^d_2}\tilde{b}_{12}\, \lambda^4& ~~e^{-i(4\theta^d_2+\theta^d_3)}\tilde{b}_{13}\,\lambda^6 \\
 \cdot & ~~b_{01} &~~~ e^{-i(7\theta^d_2+2\theta^d_3)}\tilde{b}_{23}\, \lambda^5 \\
 \cdot& ~~\cdot & b_{02}
\end{array}
\right)\ ,\\
\frac{(\tilde{m}_{u}^2)_{{RR}_{\text{GUT}}}}{m_0^2}&\approx&\left(
\begin{array}{ccc}
b_{01}&~~~~e^{-i\theta^d_2}\tilde{b}_{12}\,\lambda^4~ &~ \tilde{b}_{13}\,\lambda^6 \\
 \cdot &b_{01}&~~e^{i(5\theta^d_2+\theta^d_3)}\tilde{b}_{23}\, \lambda^5 \\
\cdot & \cdot &b_{02}
\end{array}
\right).~~~
\end{eqnarray}%


\subsubsection*{Down-type quark sector:}

\begin{eqnarray}
\frac{\tilde{A}^d_\text{GUT}}{A_0}&\approx&\left(
\begin{array}{ccc}
~\tilde{a}^d_{11}\,\lambda^6&~~~\tilde{a}^d_{12}\,\lambda^5~~~ & \tilde{a}^d_{12}\,\lambda^5 \\
-\tilde{a}^d_{12}\,\lambda^5&~~~\tilde{a}^d_{22}\,\lambda^4~~~&\tilde{a}^d_{23}\,\lambda^4\\
~e^{-i\theta^d_2}\tilde{a}^d_{31}\,\lambda^6&~~~\tilde{a}^d_{32}\,\lambda^6~~~&\tilde{a}^d_{33}\,\lambda^2
\end{array}
\right) \ ,\\
 \frac{(\tilde{m}_{d}^2)_{{LL}_{\text{GUT}}}}{m_0^2}&\approx&\left(
\begin{array}{ccc}
b_{01}~&~\tilde{B}_{12}\,\lambda^3~& ~e^{i\theta^d_2}\tilde{B}_{13}\,\lambda ^4 \\
 \cdot ~&~b_{01}~& ~~~\tilde{B}_{23}\, \lambda^2\\
\cdot ~&~ \cdot ~&~b_{02}
\end{array}
\right) \ ,\\
\frac{(\tilde{m}_{d}^2)_{{RR}_{\text{GUT}}}}{m_0^2}&\approx&\left(
\begin{array}{ccc}
 1~& ~~e^{i\theta^d_2}\tilde{R}_{12}\,\lambda^4&~-e^{i\theta^d_2}\tilde{R}_{12}\,\lambda^4   \\
\cdot~ &~1 &~~-\tilde{R}_{12}\,\lambda^4\\
 \cdot~ &~ \cdot~&~ 1 
\end{array}
\right) \ .
\end{eqnarray}%


\subsubsection*{Charged lepton sector:}

\begin{eqnarray}
\frac{\tilde{A}^e_\text{GUT}}{A_0}&\approx&\left(
\begin{array}{ccc}
\frac{1}{3}\tilde{a}^d_{11}\,\lambda^6&~~~e^{i\theta^d_2}\tilde{a}^d_{12}\,\lambda^5~~~ & \tilde{a}^d_{31}\,\lambda^6 \\
-e^{-i\theta^d_2}\tilde{a}^d_{12}\,\lambda^5&~~~~~3\tilde{a}^d_{22}\,\lambda^4~~~&\tilde{a}^e_{23}\,\lambda^6\\
-e^{-i\theta^d_2}\tilde{a}^d_{12}\,\lambda^5&~~~~~3\tilde{a}^d_{23}\,\lambda^4~~~&\tilde{a}^d_{33}\,\lambda^2
\end{array}
\right) \ , \label{AeGUT}\\
\frac{(\tilde{m}_{e}^2)_{{LL}_{\text{GUT}}}}{m_0^2}&\approx&\left(
\begin{array}{ccc}
1~&~~~\tilde{R}_{12}\,\lambda^4&~~~-\tilde{R}_{12}\,\lambda^4\\
 \cdot ~&~1~&~~~-\tilde{R}_{12}\,\lambda^4\\
 \cdot ~&~ \cdot ~&~~~1
\end{array}
\right) \ ,\\
\frac{(\tilde{m}_{e}^2)_{{RR}_{\text{GUT}}}}{m_0^2}&\approx&\left(
\begin{array}{ccc}
b_{01}~&~-e^{i\theta^d_2}\frac{1}{3}\tilde{B}_{12}\,\lambda ^3&~~ \frac{1}{3}\tilde{B}_{13}\,\lambda ^4 \\
 \cdot ~&~b_{01}~&~~3\tilde{B}_{23}\,\lambda ^2\\
 \cdot~&~ \cdot ~&~b_{02}
\end{array}
\right) \ .\label{meRRGUT}
\end{eqnarray}%




\section{Mass insertion parameters}
\cleqn

In supersymmetry, flavour changing processes are induced by the mismatch of
fermion and sfermion mass eigenstates. Having changed the basis of the
superfields to the SCKM basis, the Yukawa matrices are diagonal. Thus, the
off-diagonal entries of the scalar mass matrices determine the size of the
resulting FCNCs. As both the left- and the right-handed fermions have their
own scalar partners, there are three types of scalar mass matrices
\begin{equation}
m^2_{\tilde{f}_{LL}}\!\!=(\tilde{m}_f^2)_{LL}+\tilde{Y}_f\tilde{Y}_f^\dagger \upsilon^2_{u,d}\ ,~\quad
m^2_{\tilde{f}_{RR}}\!\!=(\tilde{m}_f^2)_{RR}+\tilde{Y}_f^\dagger\tilde{Y}_f
\upsilon^2_{u,d}\ , ~\quad
m^2_{\tilde{f}_{LR}}\!\!=\tilde{A}_f \upsilon_{u,d}-\mu \tilde{Y}_f
\upsilon_{d,u}\ ,
\label{fullmasses}
\end{equation}%
where $\mu$ is the higgsino mass which we take to be real.
In Eq.~(\ref{fullmasses}), the first contribution on the right-hand sides 
originates from the soft breaking 
Lagrangian, while the second term is the supersymmetric $F$-term contribution
to the scalar masses. We note that it is formally possible to define
$m^2_{\tilde{f}_{RL}}\equiv (m^2_{\tilde{f}_{LR}})^\dagger$.

From the model building perspective, a convenient measure of flavour violation
is provided by a set of dimensionless parameters, known as the mass insertion
parameters. These are defined as~\cite{Gabbiani:1996hi}
\begin{eqnarray}
(\delta^f_{LL})_{ij}=\frac{(m^2_{\tilde{f}_{LL}})_{ij}}{\langle
    m_{\tilde{f}}\rangle ^2_{LL}}\ ,\qquad
(\delta^f_{RR})_{ij}=\frac{(m^2_{\tilde{f}_{RR}})_{ij}}{\langle
    m_{\tilde{f}}\rangle ^2_{RR}}\ ,\qquad
(\delta^f_{LR})_{ij}=\frac{(m^2_{\tilde{f}_{LR}})_{ij}}{\langle
    m_{\tilde{f}}\rangle ^2_{LR}}\ ,
\label{ins}
\end{eqnarray}%
where the average masses in the denominators are
\begin{equation}
\langle m_{\tilde{f}}\rangle
^2_{AB}=\sqrt{(m^2_{\tilde{f}_{AA}})_{ii}(m^2_{\tilde{f}_{BB}})_{jj}}\ .\label{mav}
\end{equation}


\subsection[Mass insertion parameters at the GUT scale]{Mass insertion
  parameters $\boldsymbol{\delta}$ at the GUT scale} 

Inserting the results of Section~\ref{section:SCKM}, it is straightforward to
calculate the mass insertion parameters at the GUT scale. The full LO
expressions are given in Appendix~\ref{app:defMI}. In the following we only
report the flavour structure of the various $\delta$s in terms of their
$\lambda$-suppression. 
\begin{eqnarray}\label{eq:gutdeltau}
&&\delta^u_{LL_{\text{GUT}}}\sim\left(
\begin{array}{ccc}
1  & \lambda^4 & \lambda^6 \\
\cdot &1 &\lambda^5 \\
\cdot &\cdot  &1
\end{array}
\right),~~~
\delta^u_{RR_{\text{GUT}}}\sim\left(
\begin{array}{ccc}
1  & \lambda^4 & \lambda^6 \\
\cdot &1 &\lambda^5 \\
\cdot &\cdot  &1
\end{array}
\right),~~~
\delta^u_{LR_{\text{GUT}}}\sim\left(
\begin{array}{ccc}
\lambda^8 & 0& 0 \\
0&\lambda^4&\lambda^7 \\
0 &\lambda^7&1
\end{array}
\right),~~~~\\[-2mm]
\label{eq:gutdeltad}
&&\delta^d_{LL_{\text{GUT}}}\sim\left(
\begin{array}{ccc}
1  & \lambda^3 & \lambda^4 \\
\cdot &1 &\lambda^2 \\
\cdot &\cdot  &1
\end{array}
\right),~~~
\delta^d_{RR_{\text{GUT}}}\sim\left(
\begin{array}{ccc}
1  & \lambda^4 & \lambda^4 \\
\cdot &1 &\lambda^4 \\
\cdot &\cdot  &1
\end{array}
\right),~~~
\delta^d_{LR_{\text{GUT}}}\sim\left(
\begin{array}{ccc}
\lambda^6 &\lambda^5&\lambda^5 \\
\lambda^5&\lambda^4&\lambda^4\\
\lambda^6&\lambda^6&\lambda^2
\end{array}
\right),~~~~\\[-2mm]
\label{eq:gutdeltae}
&&\delta^e_{LL_{\text{GUT}}}\sim\left(
\begin{array}{ccc}
1  & \lambda^4 & \lambda^4 \\
\cdot &1 &\lambda^4 \\
\cdot &\cdot  &1
\end{array}
\right),~~~
\delta^e_{RR_{\text{GUT}}}\sim\left(
\begin{array}{ccc}
1  & \lambda^3 & \lambda^4 \\
\cdot &1 &\lambda^2 \\
\cdot &\cdot  &1
\end{array}
\right),~~~
\delta^e_{LR_{\text{GUT}}}\sim\left(
\begin{array}{ccc}
\lambda^6 &\lambda^5&\lambda^6 \\
\lambda^5&\lambda^4&\lambda^6\\
\lambda^5&\lambda^4&\lambda^2
\end{array}
\right).~~~~
\end{eqnarray}


\subsection[Effects of renormalisation group running]{Effects of RG running}

Having calculated the GUT scale mass insertion parameters, it is now necessary
to consider their evolution down to the electroweak scale. Only then are we
able to compare the predictions of the model to experimental measurements of
flavour observables. This evolution is described by the RG equations which are
given explicitly in Appendix~\ref{App:RGEs} in the SCKM basis. 
Technically, we perform the RG running in two stages, first from $M_\text{GUT}$ to
$M_{R}$ where the right-handed neutrinos are integrated out, and then from
$M_{R}$ to $M_\text{SUSY}\sim M_{W}$. 
In order to derive analytical results, we estimate the effects of the running using the
leading logarithmic approximation. 
As the Yukawa matrices themselves are also affected by the running, it is
necessary to apply further basis transformations on the superfields which
diagonalise the low energy Yukawas matrices.

Details of the various steps involved in calculating the low energy mass
insertion parameters can be found in Appendix~\ref{sec:defRG}. For the
down-type squarks and the charged sleptons, the resulting effects can simply be
absorbed into new order one coefficients. It is interesting to see that this is not
the case for the up-type squarks, where the order of the (13) and (23) elements
of $\delta^u_{LR}$ gets modified. For completeness, we present the flavour
structure of the low energy $\delta$s in terms of their $\lambda$-suppression,
which should be compared to Eqs.~(\ref{eq:gutdeltau}-\ref{eq:gutdeltae}).
\begin{eqnarray}\label{eq:deltau}
&&\delta^u_{LL}\sim\left(
\begin{array}{ccc}
1  & \lambda^4 & \lambda^6 \\
\cdot &1 &\lambda^5 \\
\cdot &\cdot  &1
\end{array}
\right),~~~
\delta^u_{RR}\sim\left(
\begin{array}{ccc}
1  & \lambda^4 & \lambda^6 \\
\cdot &1 &\lambda^5 \\
\cdot &\cdot  &1
\end{array}
\right),~~~
\delta^u_{LR}\sim\left(
\begin{array}{ccc}
\lambda^8 & 0& \lambda^7 \\
0&\lambda^4&\lambda^6 \\
0 &\lambda^7&1
\end{array}
\right),~~~~\\[-2mm]
\label{eq:deltad}
&&\delta^d_{LL}\sim\left(
\begin{array}{ccc}
1  & \lambda^3 & \lambda^4 \\
\cdot &1 &\lambda^2 \\
\cdot &\cdot  &1
\end{array}
\right),~~~
\delta^d_{RR}\sim\left(
\begin{array}{ccc}
1  & \lambda^4 & \lambda^4 \\
\cdot &1 &\lambda^4 \\
\cdot &\cdot  &1
\end{array}
\right),~~~
\delta^d_{LR}\sim\left(
\begin{array}{ccc}
\lambda^6 &\lambda^5&\lambda^5 \\
\lambda^5&\lambda^4&\lambda^4\\
\lambda^6&\lambda^6&\lambda^2
\end{array}
\right),~~~~\\[-2mm]
\label{eq:deltae}
&&\delta^e_{LL}\sim\left(
\begin{array}{ccc}
1  & \lambda^4 & \lambda^4 \\
\cdot &1 &\lambda^4 \\
\cdot &\cdot  &1
\end{array}
\right),~~~
\delta^e_{RR}\sim\left(
\begin{array}{ccc}
1  & \lambda^3 & \lambda^4 \\
\cdot &1 &\lambda^2 \\
\cdot &\cdot  &1
\end{array}
\right),~~~
\delta^e_{LR}\sim\left(
\begin{array}{ccc}
\lambda^6 &\lambda^5&\lambda^6 \\
\lambda^5&\lambda^4&\lambda^6\\
\lambda^5&\lambda^4&\lambda^2
\end{array}
\right).~~~~
\end{eqnarray}






\section{Conclusion}
\cleqn

Despite its tremendous success, the Standard Model of particle physics is
widely viewed as the low energy limit of a more fundamental
theory. In order to understand the nature of flavour in such extensions
of the SM it is necessary to answer the following three questions.
\newpage
\begin{enumerate}
\item Why are there three families of quarks and leptons?
\item How does the structure of fermion masses and mixing arise?
\item Why is the amount of flavour violation
   induced by new physics so small?
\end{enumerate}
 From the phenomenological point of view, the third question is usually
addressed by means of {\it ad hoc} assumptions such as e.g. Minimal Flavour
Violation, where all sources of flavour violation are intimately linked
to the
flavour structure of the Yukawa matrices. However, the concept of MFV is not
a theory of flavour as such. Moreover, it does not seem to provide a
framework in
which the first two questions of the flavour puzzle can be addressed in
a satisfactory
way.

In this paper, we have investigated the issue of flavour violation within a
supersymmetric GUT model of flavour which is based on the simple family
symmetry
$S_4\times U(1)$~\cite{M2}. The existence of three families of quarks and
leptons is related to the non-Abelian factor of the family symmetry
whose triplets are the only faithful irreducible representations. The
structure of the Yukawa matrices arises from the breaking of the family
symmetry. This aspect was thoroughly studied in~\cite{M1,M2} where it was
shown to provide a good description of all quark and lepton masses, mixings
and CP violation.

Applying the family symmetry on the soft SUSY breaking
sector, we have worked out the mass insertion parameters which describe the
sources of flavour violation beyond the SM. Our calculation relies on the
assumption that the SUSY breaking mechanism respects the family
symmetry. Working in an expansion in powers of the Wolfenstein
parameter~$\lambda$, we take into account the effect of canonical
normalisation as
well as renormalisation group running. Our results for the low energy mass
insertion parameters are summarised in
Eqs.~(\ref{eq:deltau}-\ref{eq:deltae}),
with the explicit expressions given in Appendix~\ref{App:LowMIs}.
We find that $\delta^f_{LL}$ and $\delta^f_{RR}$ are approximately equal to
the identity with only small off-diagonal entries. Considering the
parameters
$\delta^f_{LR}$ we observe that the diagonal elements feature the same
hierarchies as the corresponding diagonal Yukawa matrices $\tilde Y^f$,
while
the off-diagonal elements are strongly suppressed. This shows that our
$S_4\times U(1)$ SUSY GUT approximately reproduces the effects of low energy
MFV, where one would simply impose $\delta^f_{LL}= \delta^f_{RR} =
\mathds 1$
and $\delta^f_{LR} \propto \tilde Y^f$.
The phenomenological implications of the deviations form MFV will
be discussed quantitatively in a dedicated paper~\cite{future}, where we
will present and discuss the predictions of our model of flavour with
respect to a number of different flavour observables in detail.

\newpage




\section*{Acknowledgements}

We thank Claudia Hagedorn for helpful discussions throughout this project. 
MD and SFK acknowledge partial support from the STFC Consolidated ST/J000396/1
grant and the European Union FP7 ITN-INVISIBLES 
(Marie Curie Actions, PITN-GA-2011-289442). 
CL is supported by the Deutsche Forschungsgemeinschaft (DFG) within
the Research Unit FOR 1873 ``Quark Flavour Physics and Effective Field Theories''.




\section*{Appendix}




\begin{appendix}

\section{$\boldsymbol{S_4}$ and CP symmetry}
\label{App:S4GroupTheory}
\cleqn

The non-Abelian finite group $S_4$ can be defined in terms of the presentation
\begin{eqnarray}\nonumber
&& S^2 = \mathds{1} \; , \;\; ~~~T^3 = \mathds{1} \; , \;\; ~~~U^2= \mathds{1} \; ,\\[0mm]
\nonumber
&&  (S T)^3 = \mathds{1} \; , \;\; ~~~(S U)^2 = \mathds{1} \; , \;\; ~~~(T U)^2 = \mathds{1} \; , \;\;
  ~~~(S T U)^4 = \mathds{1} \; ,
\end{eqnarray}
where $S$, $T$ and $U$ denote the generators of the group. Explicit matrix
representations are basis dependent. In this work we apply the basis where the
$T$ generator is diagonal and complex for the doublet and triplet
representations. Defining $\omega = e^{2\pi i/3}$, we have
\begin{center}
\begin{math}
\begin{array}{llll}
{\bf 1}:       & ~~~S=1 \; , & ~~~T=1 \; ,  & ~~~U=1 \ ,\\[2mm]
{\bf 1^\prime}:    & ~~~S=1  \; , & ~~~T=1 \; ,  & ~~~U=-1\ ,\\[2mm]
{\bf 2}: & ~~~S= \left( \begin{array}{cc}
    1&0 \\
    0&1
    \end{array} \right) \; ,
    & ~~~T= \left( \begin{array}{cc}
    \omega&0 \\
    0&\omega^2
    \end{array} \right) \; ,
    & ~~~U=  \left( \begin{array}{cc}
    0&1 \\
    1&0
    \end{array} \right)\ ,\\[2mm]
{\bf 3}: & ~~~S= \frac{1}{3} \left(\begin{array}{ccc}
    -1& 2  & 2  \\
    2  & -1  & 2 \\
    2 & 2 & -1
    \end{array}\right) \; ,
    & ~~~T= \left( \begin{array}{ccc}
    1 & 0 & 0 \\
    0 & \omega^{2} & 0 \\
    0 & 0 & \omega
    \end{array}\right) \; ,
    & ~~~U= - \left( \begin{array}{ccc}
    1 & 0 & 0 \\
    0 & 0 & 1 \\
    0 & 1 & 0
    \end{array}\right)\ ,\\[2mm]
{\bf 3^\prime}: & ~~~S= \frac{1}{3} \left(\begin{array}{ccc}
    -1& 2  & 2  \\
    2  & -1  & 2 \\
    2 & 2 & -1
    \end{array}\right) \; ,
    & ~~~T= \left( \begin{array}{ccc}
    1 & 0 & 0 \\
    0 & \omega^{2} & 0 \\
    0 & 0 & \omega
    \end{array}\right) \; ,
    & ~~~U= +\left( \begin{array}{ccc}
    1 & 0 & 0 \\
    0 & 0 & 1 \\
    0 & 1 & 0
    \end{array}\right)\ .
\end{array}
\end{math}
\end{center}
The corresponding Clebsch-Gordan coefficients are all real and can be found
e.g. in~\cite{M1}. 

In addition to the flavour symmetry $S_4$, we impose the canonical CP symmetry
in our theory. As has been discussed in the literature, see
e.g.~\cite{GfCPgeneral,GrimusRebelo}, the consistent combination of a
flavour and a CP symmetry requires certain 
conditions to be fulfilled; in particular that the subsequent application of a
CP, a flavour and a further CP transformation leads to a transformation
belonging to the flavour group. The possibility to combine the group $S_4$
with CP has been explored previously, see e.g.~\cite{GfCPgeneral,S4CP}.
Here we are interested in combining $S_4$ symmetry, defined in the 
above basis, with the canonical CP transformation, i.e. the CP transformation
that acts trivially in flavour space with $X_{{\bf r}}=\mathds 1$ for all
representations ${\bf r}$ of $S_4$. Note that this particular CP
transformation $X_{{\bf r}}$ fulfils the constraints of being a unitary and
symmetric matrix. Moreover, it represents a consistent choice for a CP
transformation (see e.g.\cite{S4CP}), which corresponds to the
involutionary automorphism that maps the generators $S$, $T$ and $U$ in the
following way 
\be
S \;\;\; \rightarrow \;\;\; S \;, \;\;\;\; T \;\;\; \rightarrow \;\;\; T^2=T^{-1} \;\;\; \mbox{and} \;\;\; U \;\;\; \rightarrow \;\;\; U \; ,\label{eq:autom}
\ee
since $S$ and $U$ are represented by real matrices in our chosen basis, while
the generator $T$ is given as a diagonal complex matrix in the two- and
three-dimensional representations. 
As with all automorphisms of $S_4$, this is an inner one. In particular, one
can check that the automorphism of Eq.~\eqref{eq:autom} is
``class-inverting''~\cite{ChenRatz}, i.e. it maps the group element~$g$ into
the class which includes~$g^{-1}$. This is true, since all automorphisms are
inner ones and all classes of $S_4$ are ambivalent, i.e. the elements~$g$
and~$g^{-1}$ are in the same class. 

With only real Clebsch-Gordan coefficients, a canonical CP symmetry imposed on
the theory entails that all coefficients in the (super-)potential are real.
Moreover, we observe that the residual symmetry in the neutrino sector
at LO comprises the CP symmetry if all three neutrino flavons share the same
phase factor. Following the comments of Footnote~\ref{footnote:radicand} of
Appendix~\ref{App:AppendixA}, this is the case in our setup, cf. also
Eqs.~(\ref{correlations},\ref{eq:LOflavonphases}), so that the common phase
can be factored out of the neutrino mass matrix, leading to an effective LO
result which conserves CP. Furthermore, the canonical CP transformation
$X_{{\bf r}}=\mathds 1$ commutes with the Klein group generated by $S$
and $U$ and thus at LO the residual symmetry is given by the direct product
$Z_2 \times Z_2 \times \mbox{CP}$.




\section{Vacuum alignment} \label{App:AppendixA}
\cleqn

The vacuum alignment of the flavon fields
is achieved by coupling them to a set of so-called driving fields and requiring
the $F$-terms of the latter to vanish. These driving fields, whose
transformation properties under the family symmetry are shown in
Table~\ref{drivingfields}, are SM gauge singlets and carry a charge of $+2$
under a continuous $R$-symmetry. The flavons and the GUT Higgs fields are
uncharged under this $U (1)_R$, whereas the supermultiplets containing the SM
fermions (or right-handed neutrinos) have charge $+1$. As the superpotential
must have a $U(1)_R$ charge of $+2$, the driving fields can only appear
linearly and cannot have any direct interactions with the SM fermions or
the right-handed neutrinos. 

\begin{table}[h]
\begin{center}
$$
\begin{array}{|c||c|c|c|c|c|c|c|c|c|c|c|c|c|}\hline
\text{Field}\!\!\phantom{\Big|} &  X^d_1 & \ol{X}^d_1 & X^{\nu d}_{1'} & X^u_1 & Y^{du}_2 & Y^d_2 & Y^\nu_2 &  Z^{\nu}_{3'} & V_0 & V_1 & V_\eta & X^\text{new}_1  & \tilde{X}^\text{new}_{1'} \\\hline
\!SU(5)\!\!\!\phantom{\Big|} & \bf 1 & \bf 1 & \bf 1 & \bf 1 &\bf  1 &\bf  1 &\bf 1
&\bf 1&\bf 1&\bf 1&\bf 1&\bf 1&\bf 1\\\hline
S_4\!\!\phantom{\Big|} & \bf 1&\bf 1&\bf {1^{\prime}}&\bf 1&\bf 2&\bf 2&\bf 2&\bf {3^{\prime}}&\bf 1&\bf 1&\bf 1^{(\prime)}&\bf
1& \bf {1^{\prime}}\\\hline 
U(1)\!\!\phantom{\Big|} & -2& 14&3 & 10&9 & 6&\!-16\! &\!-16\! &0 &-8 &  -7&  18&15  \\\hline
\end{array}
$$
\end{center}
\caption{\label{drivingfields} The transformation properties of the driving
  fields, as introduced in~\cite{M2}, which serve to align the flavon VEVs.}   
\end{table}

The LO alignment of the flavon fields, see Eq.~\eqref{LeadingFlavonVevs1}, has
been thoroughly discussed in~\cite{M1,M2}. The particular setup also provides
correlations amongst the VEVs. As described in Appendix D of~\cite{M1} and in
Section~4 of~\cite{M2},\footnote{The introduction of the new flavon field 
$\eta$ in~\cite{M2} favours the exchange of the $S_4$ doublet driving field
$V_2$, which was introduced in~\cite{M1}, by the $S_4$ singlet field
$V_1$. Furthermore, the field $V_\eta$, transforming in the same
representation of $S_4$ as $\eta$,  is introduced in order to relate the
new flavon field to an explicit mass scale.}  
the vanishing of the $F$-terms of the driving fields $X^\text{new}_1$,
$\tilde{X}^\text{new}_{1'}$, $Y^{\nu}_2$, $Z^{\nu}_{3'}$, $V_0$, $V_1$ and
$V_\eta$ gives rise to the relations\footnote{\label{footnote:radicand}The
proportionality constant between $\phi^\nu_{3'}$ and $\phi^\nu_{2}$ is a
square root of an order one real number, which we assume to be positive,
such that $\phi^\nu_{3'}$ and $\phi^\nu_{2}$ have the same phases.}
\begin{eqnarray}
\nonumber
&&\phi^u_2\sim \phi^d_2\,\tilde{\phi}^d_3\ ,\qquad
\phi^\nu_1\sim\phi^\nu_2\sim \phi^\nu_{3'}\ ,\qquad
(\phi^d_3)^2\phi^\nu_i\in\mathrm{Re}\ ,\\
&&\tilde{\phi}^u_2\sim\frac{\phi^\nu_1}{\phi^\nu_2}\ ,\qquad
\tilde{\phi}^d_3\sim\phi^d_2\,(\phi^d_3)^3\ ,\qquad
\phi^\nu_{3'}\sim \frac{\eta}{(\phi^d_2)^3\phi^d_3}\ .\label{correlations}
\end{eqnarray}
Denoting the phase of each flavon VEV $\phi^f_{\rho}$ by $\theta^f_{\rho}$,
Eq.~\eqref{correlations} correlates the LO phases as\footnote{{Here and in
Eq.~\eqref{eq:phasesofshifts}, a  possible phase shift by $\pi$ has been
ignored as real coefficients can generally be positive or negative.}}  
\begin{eqnarray}
\nonumber
&&\tilde{\theta}^u_2=0\ ,\qquad
\theta^u_2=2\theta^d_2+3\theta^d_3\ ,\qquad
\tilde{\theta}^d_3=\theta^d_2+3\theta^d_3\ ,\\
&&\theta^\eta=3\theta^d_2-\theta^d_3\ ,\qquad
\theta^\nu_{3'}=\theta^\nu_2=\theta^\nu_1=-2\theta^d_3\ ,
\label{eq:LOflavonphases}
\end{eqnarray}
leaving as free variables only the two phases $\theta^d_2,~\theta^d_3$, which
correspond to the LO VEVs of the two flat superpotential directions: $\langle
\Phi^d_{2,1} \rangle$ and $\langle \Phi^d_{3,2} \rangle$ respectively.

In order to find the higher order terms of the flavon VEVs, we start by
writing each one of them as a series expansion in $\lambda$, up to and
including order $\lambda^{12}$. For example, the leading operators of the
superpotential fix $\langle\Phi^u_{2,1}\rangle/M$ to be zero up to
$\lambda^4$, while $\langle\Phi^u_{2,2}\rangle/M$ has to be
$\phi^u_2\,\lambda^4$ \cite{M1}. When considering the subleading operators,
the VEVs of $\Phi^u_{2,1}$ and $\Phi^u_{2,2}$ receive corrections (shifts)
which we parametrise as 
\begin{eqnarray}
 \frac{\langle \Phi^u_2 \rangle}{M} &=& \left(
\begin{array}{ccc}
0\\
\phi^u_2\,\lambda^4
\end{array}
\right)+\left(
\begin{array}{ccc}
\sum\limits_{n=5}^{12}\delta^u_{2,1_{(n)}} \lambda^n\\
\sum\limits_{n=5}^{12}\delta^u_{2,2_{(n)}} \lambda^n
\end{array}
\right)\label{Phiu2}.
\end{eqnarray}
All flavon VEVs are parametrised in a similar manner. The aim is to find the
order of $\lambda$ at which each shift $\delta$ has to be non-zero. The 
computation consists of taking into account all possible operators and solving
the $F$-term conditions resulting from the set of driving field order by order 
in $\lambda$, up to $\lambda^{12}$. 
Each vanishing expression is solved for the lowest order shift involved. At
the end, all shifts can be expressed in terms of the LO flavon VEVs. We find 
\begin{eqnarray}
\nonumber \frac{\langle \Phi^u_2 \rangle}{M}&\!=\!& \left(
\begin{array}{ccc}
\delta^u_{2,1}\,\lambda^8\\
\phi^u_2\,\lambda^4+\delta^u_{2,2}\,\lambda^5
\end{array}
\right)\!,~~~~
 \frac{\langle \tilde{\Phi}^u_2 \rangle}{M} \!=\! \left(
\begin{array}{ccc}
\tilde{\delta}^u_{2,1}\, \lambda^6\\
\tilde{\phi}^u_2\,\lambda^4+\tilde{\delta}^u_{2,2}\,\lambda^5
\end{array}
\right)\!,~~~~
\frac{\langle \eta \rangle}{M} = \phi^\eta\lambda^4+
\delta^\eta\lambda^5,
\\
\nonumber \frac{\langle \Phi^d_3 \rangle}{M}&\!=\!&\left(
\begin{array}{ccc}
\delta^d_{3,1}\, \lambda^6\\
\phi^d_3\,\lambda^2\\
\delta^d_{3,3} \,\lambda^6
\end{array}
\right)\!,~~~~
\frac{\langle \tilde{\Phi}^d_3 \rangle}{M}\!=\!\left(
\begin{array}{ccc}
\tilde{\delta}^d_{3,1}\, \lambda^7\\
-\left(\tilde{\phi}^d_3\,\lambda^3+\tilde{\delta}^d_{3,2_{(4)}}\,\lambda^4+\tilde{\delta}^d_{3,2_{(5)}}\lambda^5\right)\\
~~~\tilde{\phi}^d_3\,\lambda^3+\tilde{\delta}^d_{3,2_{(4)}}\,\lambda^4+\tilde{\delta}^d_{3,3_{(5)}}\lambda^5
\end{array}
\right)\!,~~~~
\frac{\langle \Phi^d_2 \rangle}{M}\!=\!\left(
\begin{array}{ccc}
\phi^d_2\,\lambda\\
\delta^d_{2,2}\, \lambda^7
\end{array}
\right)\!,\\
\frac{\langle \Phi^\nu_{3'} \rangle}{M}&\!=\!&\left(
\begin{array}{ccc}
1\\1\\1
\end{array}
\right)\!\left(\phi^\nu_{3'}\lambda^4\!+\delta^\nu_{3'} \lambda^5\right)\!,~~~
\frac{\langle \Phi^\nu_2 \rangle}{M}\!=\!\left(
\begin{array}{ccc}
\phi^\nu_2\,\lambda^4\!+\delta^\nu_{2,1}\,\lambda^5\\
\phi^\nu_2\,\lambda^4\!+\delta^\nu_{2,2}\,\lambda^5
\end{array}
\right)\!,~~~
 \frac{\langle \Phi^\nu_1 \rangle}{M}= \phi^{\nu}_1\lambda^4\!+
\delta^\nu_{1}\lambda^5\!.
\label{eq:vevs+shifts}
\end{eqnarray}
Note that the shifts presented in Eq.~\eqref{eq:vevs+shifts} are the first
non-trivial ones. However, in our calculations of the mass matrices we take
into account all shifts up to $\mathcal O(\lambda^8)$. It should be pointed
out that the alignment of $\Phi^\nu_{3'}$ is not perturbed up to order
$\lambda^8$, so that it preserves the $S$ symmetry to that level. On the
other hand, the alignment of $\Phi^\nu_{2}$ is already perturbed at order
$\lambda^5$ which, however, does not break the $S$ generator as it is nothing
but the identity for the doublet representation. Taking into account also CN
effects, one can show that $m^{\text{eff}}_{\nu}$ has the form of
Eq.~\eqref{eq:efflightnu} up to $\mathcal O(\lambda^7)$.  

Eq.~\eqref{eq:vevs+shifts} is in agreement with the discussion presented in
Section 4 of~\cite{M2}, barring the absorptions of
$\delta^u_{2,2}\,\lambda$,~$\tilde{\delta}^u_{2,2}\,\lambda$,
$\tilde{\delta}^d_{3,2_{(4)}}\,\lambda$, $\delta^\nu_{3'} \lambda$, 
$\delta^\nu_1 \lambda$ and $\delta^\eta\,\lambda$ into the corresponding LO
VEVs. Being interested in the CP transformation properties of the fields, such
absorptions must not be made in the current work, as the phases of shifts and
LO VEVs are generally different. In particular, we find the following
relations between the shifts and the LO VEVs,
\begin{eqnarray}
\nonumber  
&&
 \delta^u_{2,1}\sim(\phi^d_2)^2(\phi^d_3)^3\ ,\qquad
\delta^u_{2,2}\sim(\phi^d_2)^6(\phi^d_3)^4\ ,\qquad
\tilde{\delta}^u_{2,1}\sim\tilde{\delta}^u_{2,2}\sim(\phi^d_2)^4\phi^d_3 \ , \qquad
\delta^\eta\sim(\phi^d_2)^7 ,
\\
\nonumber &&
\delta^d_{3,1}\sim\delta^d_{3,3}\sim\phi^d_3\ ,~~\quad
\tilde{\delta}^d_{3,1}\sim\phi^d_2\,(\phi^d_3)^3\ , ~~\quad
\tilde{\delta}^d_{3,2_{(4)}}\sim (\phi^d_2)^5(\phi^d_3)^4\ ,~~\quad
\tilde{\delta}^d_{3,3_{(5)}}\!-\tilde{\delta}^d_{3,2_{(5)}}\sim
(\phi^d_2)^5 ,~\\
&&
\delta^d_{2,2}\sim(\phi^d_2)^5 \phi^d_3\ ,\qquad
\delta^\nu_{3'}\sim\delta^\nu_{2,1}\sim\delta^\nu_{2,2}\sim\delta^\nu_{1}\sim\frac{(\phi^d_2)^4}{\phi^d_3}\ .
\label{eq:shifts}
\end{eqnarray}%
Similar relations also hold for higher order shifts. Although such shifts have
to be taken into account when performing a systematical $\lambda$-expansion,
their explicit expressions are irrelevant for our phenomenological study.  

The phases of the LO shifts can be deduced straightforwardly from
Eq.~\eqref{eq:shifts}. Denoting the phase of $\delta^f_{\rho,i}$ by
$\theta^f_{\rho,i}$  we obtain
\begin{eqnarray}
\nonumber \theta^u_{2,1}&=&2\theta^d_2+3\theta^d_3\ ,~~~~
\theta^u_{2,2}=2(3\theta^d_2+2\theta^d_3)\ ,~~~~
\tilde{\theta}^u_{2,1}=\tilde{\theta}^u_{2,2}=4\theta^d_2+\theta^d_3\ ,~~~~ 
\text{arg}[\delta^\eta]=7\theta^d_2\ ,
\\
\nonumber 
\theta^d_{3,1}&=&\theta^d_{3,3}=\theta^d_3\ ,~~~~~
\tilde{\theta}^d_{3,1}=\theta^d_2+3\theta^d_3\ ,~~~~~
\tilde{\theta}^d_{3,2_{(4)}}=5\theta^d_2+4\theta^d_3\ ,~~~~~
\text{arg}[\tilde{\delta}^d_{3,3_{(5)}}\!\!-\tilde{\delta}^d_{3,2_{(5)}}]=5\theta^d_2\ ,
\\
\theta^d_{2,2}&=&5\theta^d_2+\theta^d_3\ ,~~~~~
\text{arg}[\delta^\nu_{3'}]=\theta^\nu_{2,1}=\theta^\nu_{2,2}=\text{arg}[\delta^\nu_{1}]=4\theta^d_2-\theta^d_3\ .
\label{eq:phasesofshifts}
\end{eqnarray}




\section{Basis transformations}

\cleqn


\subsection{Canonical normalisation}\label{App:Pmatrices}

In order to find the transformations which map the K\"ahler potential into its
canonical form, we express the hermitian matrix $\mathcal K_A$ as in
Eq.~\eqref{Psdef}, i.e. $P_A^\dagger P_A=\mathcal K_A$. 
Note that the matrix $P_A$ is not unique since $P_A \rightarrow Q_A P_A$ with
unitary $Q_A$ will satisfy Eq.~\eqref{Psdef} just as well. Moreover, $\mathcal
K_A$ can always be decomposed as
\be
\mathcal K_A ~=~ (Q_A^\dagger \sqrt{D_A}  Q_A) (Q_A^\dagger \sqrt{D_A}  Q_A) \ ,
\ee
where ${D_A}$ is the diagonalised form of $\mathcal K_A$. Therefore it is
sufficient to find a {\it hermitian} matrix $P_A$ which satisfies
Eq.~\eqref{Psdef}, i.e. $P_A^\dagger P_A=P_A P_A=\mathcal K_A$.
Expanding $\mathcal K_A$ and $P_A$ in powers of $\lambda$,
\begin{equation}
\mathcal K_A=\sum\limits_{n=0}^{\infty}k_n \lambda^n\,,~~~~~ 
P_A =\sum\limits_{m=0}^{\infty}p_m \lambda^m\,,
\end{equation}
with $k_n,~p_n$ being matrices, allows one to calculate $P_A$
iteratively. With $k_0=\mathds 1$, the result reads 
\begin{eqnarray}
p_0=\mathds{1}\,,\qquad
p_1=\frac{1}{2}k_1\,,\qquad
p_n=\frac{1}{2}\left(k_n-\sum\limits_{j=1}^{n-1}p_{j}p_{n-j}\right)\,.
\end{eqnarray}


\subsection{SCKM transformations}\label{App:SCMKMatrices}

The SCKM rotation matrices that diagonalise the Yukawas are found through the
singular value decomposition. 
In particular, if $Y^{f}= U^f_L\tilde{Y}^{f}_\text{diag}(U^f_R)^\dagger$, then
$U^f_L$ and  $U^f_R$ consist of the eigenvectors of $Y^{f}(Y^{f})^\dagger$  
and $(Y^{f})^\dagger Y^{f}$, respectively. 
These eigenvectors are only defined up to phase transformations
\begin{eqnarray}
U^f_L&\to& U^f_L \Omega^f_L\,,~~~~~~~~~~
\Omega^f_L=\text{diag}\left({e^{i\omega^f_{L_1}},e^{i\omega^f_{L_2}},e^{i\omega^f_{L_3}}}\right),\\
U^f_R&\to& U^f_R \Omega^f_L\Omega^f_R\,,~~~~~~
\Omega^f_R=\text{diag}\left({e^{i\omega^f_{R_1}},e^{i\omega^f_{R_2}},e^{i\omega^f_{R_3}}}\right).
\end{eqnarray}
We fix the phases of the matrices $\Omega^f_L$ by requiring that the CKM and
PMNS mixing matrices are given in the standard phase convention, 
while the phases of $\Omega^f_R$ are fixed by demanding real and positive
charged fermion masses. To LO, we find the following structure of the SCKM
transformation matrices in terms of their $\lambda$-suppression.
\begin{eqnarray}
 U^u_L&\approx&\left(
\begin{array}{ccc}
 1&~~\lambda^4 &~~\lambda^6 \\
\lambda^4  &1&~~ \lambda^5 \\
\lambda^6 &~~\lambda^5 &1
\end{array}
\right)\label{UuL},~~~~~~
U^u_R\approx\left(
\begin{array}{ccc}
1&~~ \lambda^4 &~~\lambda^6\\
\lambda^4  &1&~~ \lambda^5 \\
\lambda^6 &~~\lambda^5 &1
\end{array}
\right)\label{UuRp},\\
 U^d_L&\approx&\left(
\begin{array}{ccc}
1 &~~\lambda & ~~\lambda^3 \\
\lambda& ~~1 &~~\lambda^2 \\
\lambda^4 &~~ \lambda^2&1
\end{array}
\right),\label{UdL}~~~~~~
 U^d_R\approx\left(
\begin{array}{ccc}
1 &~~\lambda &  ~~\lambda^4 \\
\lambda&~~1&~~ \lambda^4 \\
\lambda^4&~~\lambda^4&~~1
\end{array}
\right),\label{UdRp}\\
 U^e_L&\approx&
\left(
\begin{array}{ccc}
1 &~~\lambda &  ~~\lambda^4 \\
\lambda&~~1&~~ \lambda^4 \\
\lambda^4&~~\lambda^4&~~1
\end{array}
\right),
\label{UeL}~~~~~~
U^e_R\approx
\left(
\begin{array}{ccc}
1 &~~\lambda & ~~\lambda^3 \\
\lambda& ~~1 &~~\lambda^2 \\
\lambda^4 &~~ \lambda^2&1
\end{array}
\right).
\label{UeRp}
\end{eqnarray}

With these SCKM transformations, it is straightforward to calculate the CKM
mixing to leading order, 
\begin{eqnarray}
 V_{\text{CKM}_{\text{GUT}}}&=&(U^u_L)^TU^{d*}_L\approx
\left(
\begin{array}{ccc}
 1 & ~~\frac{\tilde{x}_2}{y_s} \lambda  & ~~\frac{\tilde{x}_2}{y_b}\lambda^3\\
 -\frac{\tilde{x}_2}{y_s} \lambda  & ~~1 &~~\frac{y_s}{y_b} \lambda ^2 \\
-e^{-i\theta^d_2}\frac{\tilde{x}_2^2}{y_s\,y_b} \lambda^4  & ~~-\frac{y_s}{y_b}  \lambda ^2 & 1
\end{array}
\right).
\end{eqnarray}
The associated measure of CP violation is given by the Jarlskog invariant
$J^q_{\text{CP}_\text{GUT}}$ and can be calculated from the imaginary part of
$V_{\text{CKM}_{\text{GUT}_{21}}}
V_{\text{CKM}_{\text{GUT}_{32}}}
V^\ast_{\text{CKM}_{\text{GUT}_{22}}}
V^\ast_{\text{CKM}_{\text{GUT}_{31}}}$. The explicit result can be found in
Eq.~\eqref{JCPq}.




\section{Mass insertion parameters at the GUT scale}
\label{app:defMI}
\cleqn

In the following we present the explicit expression for the various LO mass insertion
parameters at the GUT scale whose $\lambda$-suppressions have been stated in
Eqs.~(\ref{eq:gutdeltau}-\ref{eq:gutdeltae}). Using the definitions of
Eqs.~(\ref{ins},\ref{mav}), we obtain  
%
%
\begin{equation*}
\delta^u_{LL_{\text{GUT}}}\approx\left(
\begin{array}{ccc}
 1&\frac{e^{-i\theta^d_2}\tilde{b}_{12}}{b_{01}}\,\lambda^4 & \frac{e^{-i(4\theta^d_2+\theta^d_3)}\tilde{b}_{13}}{\sqrt{b_{01}\left(b_{02}+\upsilon_u^2\,y_t^2/m_0^2\right)}}\,\lambda^6\\
 \cdot& 1 &\frac{e^{-i(7\theta^d_2+2\theta^d_3)}\tilde{b}_{23}}{\sqrt{b_{01}\left(b_{02}+\upsilon_u^2\,y_t^2/m_0^2\right)}}\,\lambda^5  \\
 \cdot& \cdot&1
\end{array}
\right),
\end{equation*}
\begin{equation}
\delta^u_{RR_{\text{GUT}}}\approx\left(
\begin{array}{ccc}
 1&\frac{e^{-i\theta^d_2}\tilde{b}_{12}}{b_{01}}\,\lambda^4 & \frac{\tilde{b}_{13}}{\sqrt{b_{01}\left(b_{02}+\upsilon_u^2\,y_t^2/m_0^2\right)}}\,\lambda^6\\
 \cdot& 1 &\frac{e^{i(5\theta^d_2+\theta^d_3)}\tilde{b}_{23}}{\sqrt{b_{01}\left(b_{02}+\upsilon_u^2\,y_t^2/m_0^2\right)}}\,\lambda^5  \\
 \cdot& \cdot&1
\end{array}
\right),
\end{equation}
\begin{equation*}
\delta^u_{LR_{\text{GUT}}}\approx\frac{\upsilon_u\,\alpha_0}{m_0}\left(
\begin{array}{ccc}
\frac{\tilde{a}^u_{11}-y_u\frac{\mu}{t_\beta A_0}}{b_{01}}\lambda^8  &0 & 0 \\
 0 &\frac{\tilde{a}^u_{22}-y_c\frac{\mu}{t_\beta A_0}}{b_{01}}\lambda^4  &\frac{e^{i\theta^d_2}\tilde{a}^u_{23}}{\sqrt{b_{01}\left(b_{02}+\upsilon_u^2\,y_t^2/m_0^2\right)}}\,\lambda^7   \\
0 & \frac{e^{i(3\theta^d_2+\theta^d_3)}\tilde{a}^u_{23}}{\sqrt{b_{01}\left(b_{02}+\upsilon_u^2\,y_t^2/m_0^2\right)}}\,\lambda^7 &
\frac{\tilde{a}^u_{33}-y_t\frac{\mu}{t_\beta A_0}}{b_{02}+\upsilon_u^2\,y_t^2/m_0^2}
\end{array}
\right),
\end{equation*}

%
%

\begin{equation}
\delta^d_{LL_\text{GUT}}\!\approx\!\left(
\begin{array}{ccc}
1 ~ &~~~\frac{\tilde{B}_{12}}{b_{01}}\, \lambda^3 &~~~\frac{e^{i\theta^d_2}\tilde{B}_{13}}{\sqrt{b_{01}\,b_{02}}} \,\lambda^4 \\
\cdot &~1 &~~~\frac{\tilde{B}_{23}}{\sqrt{b_{01}\,b_{02}}}\,\lambda^2 \\
\cdot &\cdot  &1
\end{array}
\right)\!,~~~~
\delta^d_{RR_\text{GUT}}\!\approx\!\left(
\begin{array}{ccc}
1 ~ &~e^{i\theta^d_2}\tilde{R}_{12}\, \lambda^4 &~-e^{i\theta^d_2}\tilde{R}_{12}\,\lambda^4 \\
\cdot &~1 &~-\tilde{R}_{12}\,\,\lambda^4 \\
\cdot &\cdot  &1
\end{array}
\right)\!,
\end{equation}\\[-8mm]
\begin{equation*}
\delta^d_{LR_\text{GUT}}\!\approx\!\frac{\upsilon_d\,\alpha_0}{m_0}\left(
\begin{array}{ccc}
\frac{1}{\sqrt{b_{01}}}\left(\tilde{a}^d_{11}-\frac{\mu\,t_\beta}{A_0}\frac{\tilde{x}_2^2}{y_s}\right)\lambda^6 &\frac{\tilde{a}^d_{12}}{\sqrt{b_{01}}}~ \lambda^5 &\frac{\tilde{a}^d_{12}}{\sqrt{b_{01}}} \lambda^5\\
-\frac{\tilde{a}^d_{12}}{\sqrt{b_{01}}} \lambda^5 &\frac{1}{\sqrt{b_{01}}}\left(\tilde{a}^d_{22}-\frac{\mu\,t_\beta}{A_0}y_s\right)\lambda^4 &\frac{\tilde{a}^d_{23}}{\sqrt{b_{01}}}\,\lambda^4 \\
e^{-i\theta^d_2}\frac{\tilde{a}^d_{31}}{\sqrt{b_{02}}}\,\lambda^6&\frac{\tilde{a}^d_{32}}{\sqrt{b_{02}}}\,\lambda^6 &\frac{1}{\sqrt{b_{02}}}\left(\tilde{a}^d_{33}-\frac{\mu\,t_\beta}{A_0}y_b\right)\lambda^2
\end{array}
\right)\!,
\end{equation*}%

%
%

\begin{equation}
\delta^e_{LL_\text{GUT}}\!\approx\!\left(
\begin{array}{ccc}
1 ~ &~~~\tilde{R}_{12}\, \lambda^4 &~~~-\tilde{R}_{12} \,\lambda^4 \\
\cdot &~1 &~~~-\tilde{R}_{12} \,\lambda^4 \\
\cdot &\cdot  &1
\end{array}
\right)\!,~~~~
\delta^e_{RR_\text{GUT}}\!\approx\!\left(
\begin{array}{ccc}
1 ~ &~-\frac{e^{i\theta^d_2}\tilde{B}_{12}}{3\,b_{01}}\,\lambda ^3 &~\frac{\tilde{B}_{13}}{3\sqrt{b_{01}b_{02}}}\,\lambda^4 \\
\cdot &~1 &~\frac{3\tilde{B}_{23}}{\sqrt{b_{01}b_{02}}}\,\lambda^2 \\
\cdot &\cdot  &1
\end{array}
\right)\!,
\end{equation}\\[-8mm]
\begin{equation*}
\delta^e_{LR_\text{GUT}}\!\approx\!\frac{\upsilon_d\,\alpha_0}{m_0}\left(
\begin{array}{ccc}
\frac{1}{3\sqrt{b_{01}}}\left(\tilde{a}^d_{11}-\frac{\mu\,t_\beta}{A_0}\frac{\tilde{x}_2^2}{y_s}\right)\lambda^6 &\frac{e^{i\theta^d_2}\tilde{a}^d_{12}}{\sqrt{b_{01}}}~ \lambda^5 &\frac{\tilde{a}^d_{31}}{\sqrt{b_{02}}} \lambda^6\\
-\frac{e^{-i\theta^d_2}\tilde{a}^d_{12}}{\sqrt{b_{01}}} \lambda^5 &\frac{3}{\sqrt{b_{01}}}\left(\tilde{a}^d_{22}-\frac{\mu\,t_\beta}{A_0}y_s\right)\lambda^4 &\frac{\tilde{a}^e_{23}}{\sqrt{b_{02}}}\,\lambda^6 \\
-\frac{e^{-i\theta^d_2}\tilde{a}^d_{12}}{\sqrt{b_{01}}} \lambda^5&\frac{3\tilde{a}^d_{23}}{\sqrt{b_{01}}}\,\lambda^4 &\frac{1}{\sqrt{b_{02}}}\left(\tilde{a}^d_{33}-\frac{\mu\,t_\beta}{A_0}y_b\right)\lambda^2
\end{array}
\right)\!. 
\end{equation*}
These $\delta$ parameters are expressed in terms of the coefficients of the
soft mass matrices in Eqs.~(\ref{AutGUT}-\ref{meRRGUT}), where we have
defined
\begin{eqnarray}
 \tilde{b}_{12}&=&(b_{2}-b_{01}k_{2}),~~~~
\tilde{b}_{13}=-(b_4-b_{01}k_4),~~~~
\tilde{b}_{23}=-(b_{3}-b_{01}k_{3}),\label{Bt}\\
\tilde{B}_{12}&=&2\frac{\tilde{x}_2}{y_s}(b_{1}-b_{01}k_{1}),~~\:
\tilde{B}_{13}=\frac{\tilde{x}_2^2}{y_b\,y_s}(b_{01}-b_{02}),~~\:
\tilde{B}_{23}=\frac{y_s}{y_b}(b_{01}-b_{02}),~~\:
\tilde{R}_{12}=B_3-K_3,\nonumber
\end{eqnarray}
and
\begin{eqnarray}
\nonumber 
\tilde{a}^u_{11}&=&a_ue^{i(\theta^a_u-\theta^y_u)},~~~~
\tilde{a}^u_{22}=a_ce^{i(\theta^a_c-\theta^y_u)},~~~~
\tilde{a}^u_{33}=a_t,~~~~
\tilde{a}^u_{23}=z^u_2\left(\frac{a_t}{y_t}-e^{i(\theta^{z_{u_a}}_2-\theta^{z_u}_2)}\frac{z^{u_a}_2}{z^u_2}\right),\\
\nonumber \tilde{a}^d_{11}&=&\frac{\tilde{x}_2^2}{y_s}\left(2\frac{\tilde{x}^a_2}{\tilde{x}_2}e^{i(\theta^{\tilde{x}_a}_2-\theta^{\tilde{x}}_2)}-\frac{a_s}{y_s}e^{i(\theta^a_s-\theta^y_s)}\right),~~~~
\tilde{a}^d_{22}=a_se^{i(\theta^a_s-\theta^y_s)},~~~~
\tilde{a}^d_{33}=a_be^{i(\theta^a_b-\theta^y_b)},\\
\nonumber \tilde{a}^d_{12}&=&\tilde{x}_2\left(\frac{\tilde{x}^a_2}{\tilde{x}_2}
e^{i(\theta^{\tilde{x}_a}_2-\theta^{\tilde{x}}_2)}-\frac{a_s}{y_s}e^{i(\theta^a_s-\theta^y_s)}\right),~~~~
\tilde{a}^d_{23}=y_s\left(\frac{a_s}{y_s}e^{i(\theta^a_s-\theta^y_s)}-\frac{a_b}{y_b}
e^{i(\theta^a_b-\theta^y_b)}\right),
\end{eqnarray}
\begin{eqnarray}
\nonumber 
\tilde{a}^d_{31}&=&z^d_3
\left(\frac{a_b}{y_b}e^{i(\theta^a_b-\theta^y_b)}-\frac{z^{d_a}_3}{z^{d}_3}e^{i(\theta^{z_{d_a}}_3-\theta^{z_d}_3)}\right),~~~~\\
\nonumber \tilde{a}^d_{32}&=&\frac{y_s^2}{y_b}\left(\frac{a_s}{y_s}e^{i(\theta^a_s-\theta^y_s)}-\frac{a_b}{y_b}e^{i(\theta^a_b-\theta^y_b)}\right)
+z^d_2\left(\frac{a_b}{y_b}e^{i(\theta^a_b-\theta^y_b)}-\frac{z^{d_a}_2}{z^d_2}e^{i(\theta^{z_{d_a}}_2-\theta^{z_d}_2)}\right),~~~~\\
\tilde{a}^e_{23}&=&9\frac{y_s^2}{y_b}\left(\frac{a_s}{y_s}e^{i(\theta^a_s-\theta^y_s)}-\frac{a_b}{y_b}e^{i(\theta^a_b-\theta^y_b)}\right)+z^d_2
\left(\frac{a_b}{y_b}e^{i(\theta^a_b-\theta^y_b)}-\frac{z^{d_a}_2}{z^{d}_2}e^{i(\theta^{z_{d_a}}_2-\theta^{z_d}_2)}
\right).~~~~ \label{adt}
\end{eqnarray}
The phases $\theta^y_{u,c,s,b},\theta^{z_{u,d}}_i,\theta^{\tilde{x}}_2$ can be
expressed in terms of the flavon phases $\theta^d_2,\theta^d_3$ according to
Eqs.~(\ref{upphases},\ref{downphases}). This has been done in Eq.~\eqref{Bt},
but we refrain from doing so in Eq.~\eqref{adt} in order to highlight the
fact that all $\tilde{a}^f_{ij}$ become real in the limit where the contributions of
the auxiliary components of the flavon superfields to the $A$-terms
are neglected such that the relation $\theta^{a}_f=\theta^y_f$ holds.




\section{Renormalisation group equations in SCKM basis}
\label{App:RGEs}
\cleqn

The renormalisation group equations for the parameters of the superpotential
as well as the soft breaking terms are usually given in the gauge flavour
basis, see e.g.~\cite{Chung:2003fi}, with the transformation to the SCKM basis being
defined only at the electroweak scale. As already discussed in
Section~\ref{section:SCKM}, we find it useful to diagonalise the Yukawa
matrices already at the high scale. In such a high scale SCKM basis, the RGEs will
explicitly depend on the CKM mixing matrix. Here we define for convenience
\begin{eqnarray}
V&=&(U^d_L)^\dagger U^u_L=V_{\text{CKM}_\text{GUT}}^T.
\end{eqnarray}
Introducing the parameter $t=\ln(\mu/M_{x})$, with $\mu$ being the
renormalisation scale and $M_x$ the high energy scale, we have for the Yukawas
and the trilinear $A$-parameters,
{\small{
\begin{eqnarray}
\nonumber16\pi^2\frac{d\tilde{Y}^u}{dt}&=&\left(3\tilde{Y}^u\tilde{Y}^{u\dagger}+V^\dagger\tilde{Y}^d\tilde{Y}^{d\dagger}V-
\frac{16}{3}g_3^2-3g_2^2-\frac{13}{15}g_1^2+3\text{Tr}[\tilde Y^{u\dagger}\tilde Y^u]+\text{Tr}[\tilde Y^{\nu\dagger}\tilde Y^\nu]\right)\tilde Y^u,\\
\nonumber16\pi^2\frac{d\tilde{Y}^d}{dt}&=&\left(3\tilde{Y}^d\tilde{Y}^{d\dagger}+V\tilde{Y}^u\tilde{Y}^{u\dagger}V^\dagger-
\frac{16}{3}g_3^2-3g_2^2-\frac{7}{15}g_1^2+3\text{Tr}[\tilde Y^{d\dagger}\tilde Y^d]+\text{Tr}[\tilde Y^{e\dagger}\tilde Y^e]\right)\tilde Y^d,\\
16\pi^2\frac{d\tilde{Y}^e}{dt}&=&\left(3\tilde{Y}^e\tilde{Y}^{e\dagger}+U_L^{e\dagger}Y^\nu Y^{\nu\dagger}U^e_L-
3g_2^2-\frac{9}{5}g_1^2+3\text{Tr}[\tilde Y^{d\dagger}\tilde Y^d]+\text{Tr}[\tilde Y^{e\dagger}\tilde Y^e]\right)\tilde Y^e,\label{YRG}
\end{eqnarray}
}}

{\small{
\begin{eqnarray}
\nonumber&&16\pi^2\frac{d\tilde{A}^u}{dt}=\left(5\tilde{Y}^u\tilde{Y}^{u\dagger}+V^\dagger\tilde{Y}^d\tilde{Y}^{d\dagger}V
-\frac{16}{3}g_3^2-3g_2^2-\frac{13}{15}g_1^2+3\text{Tr}[\tilde Y^{u\dagger}\tilde Y^u]+\text{Tr}[Y^{\nu\dagger}Y^{\nu}]\right)\tilde{A}^u+\\[-2mm]
\nonumber&&~+
\bigg(4\tilde{A}^u\tilde{Y}^{u\dagger}+2V^\dagger
\tilde{A}^d\tilde{Y}^{d\dagger}V+\frac{32}{3}g_3^2M_3+6g_2^2M_2+\frac{26}{15}g_1^2M_1+6\text{Tr}[\tilde Y^{u\dagger}\tilde A^u]+2\text{Tr}[Y^{\nu\dagger}A^\nu]\bigg)\tilde{Y}^u,\\
\nonumber&&16\pi^2\frac{d\tilde{A}^d}{dt}=\left(5\tilde{Y}^d\tilde{Y}^{d\dagger}+V\tilde{Y}^u\tilde{Y}^{u\dagger}V^\dagger
-\frac{16}{3}g_3^2-3g_2^2-\frac{7}{15}g_1^2+3\text{Tr}[\tilde Y^{d\dagger}\tilde Y^d]+\text{Tr}[\tilde Y^{e\dagger}\tilde Y^{e}]\right)\tilde{A}^d+\\[-2mm]
\nonumber&&~+\left(4\tilde{A}^d\tilde{Y}^{d\dagger}+2V \tilde{A}^u\tilde{Y}^{u\dagger}V^\dagger+\frac{32}{3}g_3^2M_3+6g_2^2M_2+\frac{14}{15}g_1^2M_1+6\text{Tr}[\tilde Y^{d\dagger}\tilde A^d]+2\text{Tr}[\tilde Y^{e\dagger}\tilde A^e]\right)\tilde{Y}^d,\\
\nonumber&&16\pi^2\frac{d\tilde{A}^e}{dt}=\left(5\tilde{Y}^e\tilde{Y}^{e\dagger}+U^{e\dagger}_LY^\nu Y^{\nu\dagger}U^e_L
-3g_2^2-\frac{9}{5}g_1^2+3\text{Tr}[\tilde Y^{d\dagger}\tilde Y^d]+\text{Tr}[\tilde Y^{e\dagger}\tilde Y^{e}]\right)\tilde{A}^e+\\[-2mm]
&&~+\left(4\tilde{A}^e\tilde{Y}^{e\dagger}+2U^{e\dagger}_L A^\nu Y^{\nu\dagger}U^e_L+6g_2^2M_2+\frac{18}{5}g_1^2M_1+6\text{Tr}[\tilde Y^{d\dagger}\tilde A^d]+2\text{Tr}[\tilde Y^{e\dagger}\tilde A^e]\right)\tilde{Y}^e.
\end{eqnarray}%
}}%
\noindent The running of the soft scalar masses in the SCKM basis is given by
\begin{eqnarray}
\nonumber16\pi^2\frac{d}{dt}(\tilde{m}^2_u)_{LL}&=&G_Q\,\mathds
1+F^u_Q+V^\dagger F^d_QV,~~~~\\
\nonumber16\pi^2\frac{d}{dt}(\tilde{m}^2_d)_{LL}&=&G_Q\,\mathds 1+V F^u_QV^\dagger+F^d_Q,\\
\nonumber16\pi^2\frac{d}{dt}(\tilde{m}^2_e)_{LL}&=&G_L\,\mathds 1+F^e_L+F^\nu_L,\\
16\pi^2\frac{d}{dt}(\tilde{m}^2_f)_{RR}&=&G_f\,\mathds 1+F_f,~~~~~~f=u,d,e,
\end{eqnarray}
with
\begin{eqnarray}
\nonumber F^u_Q&=&\tilde{Y}^u\tilde{Y}^{u\dagger}(\tilde{m}^2_u)_{LL}+(\tilde{m}^2_u)_{LL}\tilde{Y}^u\tilde{Y}^{u\dagger}+
2\tilde{Y}^u(\tilde{m}^2_u)_{RR}\tilde{Y}^{u\dagger}+2(m_{H_u}^2)\tilde{Y}^u\tilde{Y}^{u\dagger}+2\tilde{A}^u\tilde{A}^{u\dagger},\\
\nonumber F^d_Q&=&\tilde{Y}^d\tilde{Y}^{d\dagger}(\tilde{m}^2_d)_{LL}+(\tilde{m}^2_d)_{LL}\tilde{Y}^d\tilde{Y}^{d\dagger}+
2\tilde{Y}^d(\tilde{m}^2_d)_{RR}\tilde{Y}^{d\dagger}+2(m_{H_d}^2)\tilde{Y}^d\tilde{Y}^{d\dagger}+2\tilde{A}^d\tilde{A}^{d\dagger},\\
\nonumber F^e_L&=&\tilde{Y}^e\tilde{Y}^{e\dagger}(\tilde{m}^2_e)_{LL}+(\tilde{m}^2_e)_{LL}\tilde{Y}^e\tilde{Y}^{e\dagger}+
2\tilde{Y}^e(\tilde{m}^2_e)_{RR}\tilde{Y}^{e\dagger}+2(m_{H_d}^2)\tilde{Y}^e\tilde{Y}^{e\dagger}+2\tilde{A}^e\tilde{A}^{e\dagger},\\
\nonumber F^\nu_L&=&U^{e\dagger}_LY^\nu Y^{\nu\dagger} U^e_L(\tilde{m}^2_e)_{LL}+(\tilde{m}^2_e)_{LL}U^{e\dagger}_LY^\nu Y^{\nu\dagger} U^e_L+
2U^{e\dagger}_LY^\nu m_{{N}}^2Y^{\nu\dagger} U^e_L+\\
\nonumber  &&~+2(m_{H_u}^2)U^{e\dagger}_LY^\nu Y^{\nu\dagger}
U^e_L+2U^{e\dagger}_LA^\nu A^{\nu\dagger} U^e_L \,,
\end{eqnarray}

\begin{eqnarray}
\nonumber F_u&=&2\left(\tilde{Y}^{u\dagger}\tilde{Y}^u(\tilde{m}^2_u)_{RR}+(\tilde{m}^2_u)_{RR}\tilde{Y}^{u\dagger}\tilde{Y}^u+
2\tilde{Y}^{u\dagger}(\tilde{m}^2_u)_{LL}\tilde{Y}^u+2(m_{H_u}^2)\tilde{Y}^{u\dagger}\tilde{Y}^u+2\tilde{A}^{u\dagger}\tilde{A}^u\right),\\
\nonumber F_d&=&2\left(\tilde{Y}^{d\dagger}\tilde{Y}^d(\tilde{m}^2_d)_{RR}+(\tilde{m}^2_d)_{RR}\tilde{Y}^{d\dagger}\tilde{Y}^d+
2\tilde{Y}^{d\dagger}(\tilde{m}^2_d)_{LL}\tilde{Y}^d+2(m_{H_d}^2)\tilde{Y}^{d\dagger}\tilde{Y}^d+2\tilde{A}^{d\dagger}\tilde{A}^d\right),\\
\nonumber F_e&=&2\left(\tilde{Y}^{e\dagger}\tilde{Y}^e(\tilde{m}^2_e)_{RR}+(\tilde{m}^2_e)_{RR}\tilde{Y}^{e\dagger}\tilde{Y}^e+
2\tilde{Y}^{e\dagger}(\tilde{m}^2_e)_{LL}\tilde{Y}^e+2(m_{H_d}^2)\tilde{Y}^{e\dagger}\tilde{Y}^e+2\tilde{A}^{e\dagger}\tilde{A}^e\right),~~~~~~
\end{eqnarray}

\begin{eqnarray}
\nonumber
G_Q&=&-4\left(\frac{8}{3}g_3^2|M_3|^2+\frac{3}{2}g_2^2|M_2|^2+\frac{1}{30}g_1^2|M_1|^2-\frac{1}{10}g_1^2
(m^2_{H_u}-m^2_{H_d})
\right),\\
\nonumber
G_L&=&-4\left(\frac{3}{2}g_2^2|M_2|^2+\frac{3}{10}g_1^2|M_1|^2+\frac{3}{10}g_1^2
(m^2_{H_u}-m^2_{H_d})
\right),\\
\nonumber
G_u&=&-4\left(\frac{8}{3}g_3^2|M_3|^2+\frac{8}{15}g_1^2|M_1|^2+\frac{2}{5}g_1^2
(m^2_{H_u}-m^2_{H_d})
\right),\\
\nonumber
G_d&=&-4\left(\frac{8}{3}g_3^2|M_3|^2+\frac{2}{15}g_1^2|M_1|^2-\frac{1}{5}g_1^2
(m^2_{H_u}-m^2_{H_d})
\right),\\
\nonumber  G_e&=&-4\left(\frac{6}{5}g_1^2|M_1|^2-\frac{3}{5}g_1^2
(m^2_{H_u}-m^2_{H_d})
\right).
\end{eqnarray}

For completeness, we also show the evolution of the 
$\mu$ parameter, i.e. the coupling of the bilinear superpotential term $H_u H_d$, 
\begin{eqnarray}
16\pi^2\frac{d\mu}{dt}&\!=\!&\left(\!3\text{Tr}[\tilde Y^{u\dagger}\tilde Y^u]+3\text{Tr}[\tilde Y^{d\dagger}\tilde Y^d]+\text{Tr}[\tilde Y^{e\dagger}\tilde  Y^e]+\text{Tr}[  Y^{\nu\dagger}  Y^\nu]-3g_2^2-\frac{3}{5}g_1^2\right)\!\mu\,,~~~~~
\end{eqnarray}
where $g_i,M_i,~i=1,2,3$ are the gaugino couplings and masses respectively.


\section{Renormalisation group running}
\label{sec:defRG}
\cleqn

In this appendix, we provide analytical expression for the RG
evolved Yukawa couplings, soft terms and mass insertion parameters. We
estimate the effects of RG running using the leading logarithmic
approximation. In order to formulate the two-stage running ($i$)~from
$M_{\text{GUT}}$ to $M_R$, where the right-handed neutrinos are integrated
out, and ($ii$)~from $M_R$ to $M_{\text{SUSY}}\sim M_\text{W}\equiv
M_{\text{low}}$, we introduce the parameters
\begin{eqnarray}
\eta&=&\frac{1}{16\pi^2}\ln\left(\frac{M_{\text{GUT}}}{M_{\text{low}}}\right),~~~
\eta_N=\frac{1}{16\pi^2}\ln\left(\frac{M_{\text{GUT}}}{M_{\text{R}}}\right).
\label{etas}
\end{eqnarray}
For $M_{\text{GUT}}\approx 2\times 10^{16}$ GeV, $M_{\text{R}}\approx  10^{14}$ GeV and $M_{\text{low}}\approx 10^{3}$ GeV, $\eta\approx0.19$ is of the order of our expansion parameter $\lambda\approx0.22$  and  $\eta_N\approx0.03$.

\subsection{Low energy Yukawas}

The SCKM transformations, discussed in Section~\ref{section:SCKM}, diagonalise
the Yukawa matrices at high scales. RG running to low energies re-introduces
off-diagonal elements in the low energy Yukawa matrices. 
These off-diagonal entries in $\tilde{Y}^u_{\text{low}}$ and
$\tilde{Y}^d_{\text{low}}$ are proportional to the quark masses and the
$V_\text{CKM}$ elements. As the CKM matrix features only a mild running, the
RG corrections can be treated as a perturbation. 
In $\tilde{Y}^e_{\text{low}}$, the off-diagonal terms are proportional to the
charged lepton masses and the elements of $Y^\nu$. The corresponding RG
equations are provided explicitly in Eq.~\eqref{YRG} for convenience. To LO in
$\lambda$, we find, 
\begin{eqnarray}
\tilde{Y}^u_{{\text{low}}}&\approx&\left(
\begin{array}{ccc}
1+R^y_u&0&0\\
0&1+R^y_u&0\\
0&0&1+R^y_t
\end{array}
\right)\tilde{Y}^u_{{\text{GUT}}}
-\eta\,y_b\,y_t\left(
\begin{array}{ccc}
0~~~&0~~&\tilde{x}_2\,\lambda^7\\
0~~~&0~~&y_s\,\lambda^6\\
0~~~&0~~&0
\end{array}
\right),\label{YuLow}\\
\tilde{Y}^d_{\text{low}}&{\approx}&\left(
\begin{array}{ccc}
1+R^y_d&0&0\\
0&1+R^y_d&0\\
0&0&1+R^y_b
\end{array}
\right)\tilde{Y}^d_{\text{GUT}}
+\eta\,y^2_t\left(
\begin{array}{ccc}
0~~~&0~&e^{i\theta^d_2}\frac{{\tilde{x}_2}^2}{y_s}\lambda^6\\
0~~~&0~&y_s\,\lambda^4\\
0~~~&\frac{y^2_s}{y_b}\lambda^6&0
\end{array}
\right),~~~~\\
\tilde{Y}^e_{\text{low}}&\approx&\left(
\begin{array}{ccc}
1+R^y_e&0&0\\
0&1+R^y_e&0\\
0&0&1+R^y_e
\end{array}
\right)\tilde{Y}^e_{\text{GUT}}+\eta_N\,y_D\,R_\nu\left(
\begin{array}{ccc}
0~&-3\,y_s\,\lambda^8&y_b\,\lambda^6\\
0~&0&y_b\,\lambda^6\\
0~&0&0
\end{array}
\right)\label{YeLow},
\end{eqnarray}
with
\begin{eqnarray}
  R^y_u&=&\eta\left(\frac{46}{5}g_U^2-3y_t^2\right)-3\eta_N\,y_D^2,~~~~~~
R^y_t=R^y_u-3\,\eta \,y_t^2,\\
R^y_d&=&\eta\frac{44}{5}g_U^2,~~~~~~
R^y_b=R^y_d-\eta\,y_t^2,
\\
R^y_e&=&\eta\frac{24}{5}g_U^2-\eta_N\,y_D^2,~~~~~~
R_\nu
=z^D_1-y_D(K_3+K^N_3).~~~~
\label{Rys}
\end{eqnarray}
where $g_U\approx \sqrt{0.52}$ is the universal gauge coupling constant at the
GUT scale.

\subsection{Low energy soft terms}

Similar to the Yukawa matrices, the parameters of the soft terms have to be
run down to low energies. Moreover, it is mandatory to perform further
transformations to the ``new'' SCKM basis which render
$\tilde{Y}^f_{\text{low}}$ diagonal again.
The running of the trilinear terms is similar to the one of the corresponding
Yukawas. To LO in $\lambda$, $\eta$ and $\eta_N$, we derive the following
expressions in the ``new'' SCKM basis.
\begin{eqnarray}\label{AuLow}
 \frac{\tilde{A}^u_{\text{low}}}{A_0}&\approx&\left(
\begin{array}{ccc}
1+R^y_u&0&0\\
0&1+R^y_u&0\\
0&0&1+R^y_t
\end{array}
\right)\frac{\tilde{A}^u_{\text{GUT}}}{A_0}-2\left(
\begin{array}{ccc}
R^a_u&0&0\\
0&R^a_u&0\\
0&0&R^a_t
\end{array}
\right)\tilde{Y}^u_{\text{GUT}}\\
\nonumber&&~-2\eta\,y_t\left(
\begin{array}{ccc}
0~~&0&y_b\, \tilde{x}^a_2\,e^{i(\theta^{\tilde{x}^a}_2-\theta^{\tilde{x}}_2)}\,\lambda^7\\
0~~&0&y_b\, a_s\,e^{i(\theta^a_s-\theta^y_s)}  \,\lambda^6\\
0~~&y_te^{i(3\theta^d_2+\theta^d_3)}\tilde{a}^u_{23}\lambda^7&0
\end{array}
\right),\\
 \frac{\tilde{A}^d_{\text{low}}}{A_0}&{\approx}&\left(
\begin{array}{ccc}
1+R^y_d&0&0\\
0&1+R^y_d&0\\
0&0&1+R^y_b
\end{array}
\right)\frac{\tilde{A}^d_{\text{GUT}}}{A_0}
-2\left(
\begin{array}{ccc}
R^a_d&0&0\\
0&R^a_d&0\\
0&0&R^a_b
\end{array}
\right)\tilde{Y}^d_{\text{GUT}}
\\
\nonumber &&~+
2\eta\,y_s\,y_t\left(
\begin{array}{ccc}
0&0&0 \\
0&0&a_t \,\lambda^4\\
0&\frac{1}{y_b}\left( y_s\, a_t- y_t\,\tilde{a}^d_{23}\right)\lambda^6&0
\end{array}
\right),\\
\label{AeLow}
 \frac{\tilde{A}^e_{\text{low}}}{A_0}&\approx&\left(
\begin{array}{ccc}
1+R^y_e&0&0\\
0&1+R^y_e&0\\
0&0&1+R^y_e
\end{array}
\right)
\frac{\tilde{A}^e_{\text{GUT}}}{A_0}-2R^a_e\,\tilde{Y}^e_{\text{GUT}}\\
\nonumber&&~+2\eta_N\,y_D\,R_\nu\,y_b
\left(
\begin{array}{ccc}
0~~~&0~&\frac{\alpha_D}{y_D}\,\lambda^6 \\
0~~~&0~&\frac{R^a_\nu}{R_\nu}\,\lambda^6 \\
0~~~&0~&0
\end{array}
\right),
\end{eqnarray}
with
\begin{eqnarray}
 R^a_u&=&\eta\left(\frac{46}{5}g_U^2\frac{M_{1/2}}{A_0}+3a_t\,y_t\right)
+3\eta_N\,y_D\,\alpha_D,~~~~~~
R^a_t=R^a_u+3\,\eta\,a_t\,y_t,~~~~\\
R^a_d&=&\eta\, \frac{44}{5}g_U^2\frac{M_{1/2}}{A_0},~~~~~~
R^a_b=R^a_d+\eta\,a_t\,y_t,~~~~\\
R^a_e&=&\eta\frac{24}{5}g_U^2\frac{M_{1/2}}{A_0}+\eta_N\,y_D\alpha_D,~~~~~~\\
R^a_\nu&=&
z^{D_a}_1e^{i\theta^{z_{D_a}}_1}-\alpha_D(K_3+K^N_3).
\end{eqnarray}
The first terms in Eqs.~(\ref{AuLow}-\ref{AeLow}) are analogous to the first
terms in Eqs.~(\ref{YuLow} - \ref{YeLow}); they are usually ignored.
The second terms contain the universal gaugino mass $M_{1/2}$ contributions,
which generate non-zero diagonal trilinear couplings through the running, even
for $A_0\rightarrow 0$. The sources of the off-diagonal entries in the Yukawa
couplings are also present for the trilinear soft terms. We see that the (13)
element in $\tilde{A}^u_{\text{low}}$, which was zero in $\tilde{A}^u_{\text{GUT}}$,
is now filled in, and there is an $\mathcal O(\lambda^6)$ contribution (but
additionally suppressed by a factor of $\eta$) to the (23) element, which was
of order $\lambda^7$ in $\tilde{A}^u_{\text{GUT}}$.
The (32) element in $\tilde{A}^u_{\text{low}}$, with $\tilde{a}^u_{23}$ 
given in Eq.~\eqref{adt}, is of the same order in $\lambda$ as the one that is
already present in $\tilde{A}^u_{\text{GUT}}$.  
All the off-diagonal elements generated by the running in
$\tilde{A}^d_{\text{low}}$ and in $\tilde{A}^e_{\text{low}}$ 
are of the same order in $\lambda$ as the ones that were already present at
the high scale.

Analogously to the trilinear $A$-terms, we find for the soft scalar mass, 
\begin{eqnarray}
\frac{(\tilde{m}_u^2)_{{LL}_{\text{low}}}}{m_0^2}&\approx&
\frac{(\tilde{m}_u^2)_{{LL}_{\text{GUT}}}}{m_0^2}+(6.5\,x+T^u_L)\,\mathds 1
-\eta\left(
\begin{array}{ccc}
0~&~~~0&~~y_t^2\,\frac{(\tilde{m}_u^2)_{{LL}_{\text{GUT}_{13}}}}{m_0^2}\\
\cdot &~~~0&~~y_t^2\,\frac{(\tilde{m}_u^2)_{{LL}_{\text{GUT}_{23}}}}{m_0^2}\\
\cdot &~~~\cdot &~~2R_q
\end{array}
\right)\!,
\label{muLLLow}\\
\frac{(\tilde{m}_u^2)_{{RR}_{\text{low}}}}{m_0^2}&\approx&
\frac{(\tilde{m}_u^2)_{{RR}_{\text{GUT}}}}{m_0^2}+(6.15\,x+T^u_R)\,\mathds 1
-2\eta
\left(
\begin{array}{ccc}
0&~~~0&~~y_t^2\,\frac{(\tilde{m}_u^2)_{{RR}_{\text{GUT}_{13}}}}{m_0^2}\\
\cdot &~~~0&~~y_t^2\,\frac{(\tilde{m}_u^2)_{{RR}_{\text{GUT}_{23}}}}{m_0^2}\\
\cdot &~~~\cdot &~~2R_q
\end{array}
\right)\!,~~~~~
\label{muRRLow}
\end{eqnarray}

\begin{eqnarray}
\frac{(\tilde{m}_d^2)_{{LL}_{\text{low}}}}{m_0^2}&\approx&
\frac{(\tilde{m}_u^2)_{{LL}_{\text{GUT}}}}{m_0^2}+(6.5\,x+T^d_L)\,\mathds 1
+\eta\left(
\begin{array}{ccc}
0&~0& \left(\frac{2R_q}{b_{01}-b_{02}}+y_t^2\right)\frac{(\tilde{m}_d^2)_{{LL}_{\text{GUT}_{13}}}}{m_0^2}\\
\cdot &~0&\left(\frac{2R_q}{b_{01}-b_{02}}+y_t^2\right)\frac{(\tilde{m}_d^2)_{{LL}_{\text{GUT}_{23}}}}{m_0^2}\\
\cdot &\cdot &~~-2R_q
\end{array}
\right)\!,~~~~~~~~
\label{mdLLLow}\\
\frac{(\tilde{m}_d^2)_{{RR}_{\text{low}}}}{m_0^2}&\approx&
\frac{(\tilde{m}_d^2)_{{RR}_{\text{GUT}}}}{m_0^2}+(6.1\,x+T^d_R)\,\mathds 1\,,
\label{mdRRLow}\\
\frac{(\tilde{m}_e^2)_{{LL}_{\text{low}}}}{m_0^2}&\approx& 
\frac{(\tilde{m}_e^2)_{{LL}_{\text{GUT}}}}{m_0^2}+(0.5\,x+T^e_L-2\eta_N\,R_l)\,\mathds 1-2\eta_N
\left(
\begin{array}{ccc}
0~~&~\tilde{E}_{12}&~-\tilde{E}^*_{12}\\
\cdot &0&~-\tilde{E}_{12}\\
\cdot &\cdot &0
\end{array}
\right)\!\lambda^4 ,~~~
\label{meLLLow}\\
\frac{(\tilde{m}_e^2)_{{RR}_{\text{low}}}}{m_0^2}
&\approx&\frac{(\tilde{m}_e^2)_{{RR}_{\text{GUT}}}}{m_0^2}
+(0.15\,x+T^e_R)\,\mathds 1,~~~\label{meRRLow}
\end{eqnarray}
where we have introduced the ratio $x=M_{1/2}^2/m_0^2$ and
\begin{eqnarray}
R_q&=&(2b_{02}+c_{H_u})\,y_t^2+\alpha_0^2\,a_t^2\,,\label{Rq_x}\\
\tilde{E}_{12}&=&y_D^2\left(\tilde R_{12}+B^N_3-K^N_3B^N_0\right)+R_l'-(K_3+K^N_3)R_l\,,\\
R_l&=&(1+B^N_0+c_{H_u})y_D^2+\alpha_0^2\alpha_D^2\,,\\
R_l'&=&(1+B^N_0+c_{H_u})y_D\,z^D_1+\alpha_0^2\alpha_D\,z^{D_a}_1e^{i\theta^{z_{D_a}}_1}\,,
\label{Rls}
\end{eqnarray}
with $c_{H_u} = m^2_{{H_u}_\text{GUT}} /m^2_0$. Furthermore, the small
quantities $T^f_{L,R}$ are defined as 
\begin{eqnarray}
T^u_L&=&\frac{1}{m_0^2}\left(\frac{1}{20} T+\Delta^u_L\right)\,,~~~~~~~~T^u_R=\frac{1}{m_0^2}\left(-\frac{1}{5}T+\Delta^u_R\right)\,,\\
T^d_L&=&\frac{1}{m_0^2}\left(\frac{1}{20}T+\Delta^d_L\right)\,,~~~~~~~~T^d_R=\frac{1}{m_0^2}\left(\frac{1}{5}T+\Delta^d_R\right)\,,\\
T^e_L&=&\frac{1}{m_0^2}\left(-\frac{3}{20}T+\Delta^e_L\right)\,,~~~~\,~T^e_R=\frac{1}{m_0^2}\left(\frac{3}{10}T+\Delta^e_R\right)\,,
\end{eqnarray}
with 
$T=\frac{1}{4\pi^2}
\int\limits_{\text{ln}(M_{\text{GUT}})}^{\text{ln}(M_{\text{low}})}g_U^2\,(m^2_{H_u}-m^2_{H_d})$, as well as
\begin{eqnarray}
\Delta^u_L&=&\left(\frac{1}{2}-\frac{2}{3}\sin^2(\theta_W)\right)\cos(2\beta)M_Z^2\,,~~~~~~\:\:\Delta^u_R=\frac{2}{3}\sin^2(\theta_W)\cos(2\beta)M_Z^2\,,\\
\Delta^d_L&=&\left(-\frac{1}{2}+\frac{1}{3}\sin^2(\theta_W)\right)\cos(2\beta)M_Z^2\,,~~~~~\Delta^d_R=-\frac{1}{3}\sin^2(\theta_W)\cos(2\beta)M_Z^2\,,~~~~~~\\
\Delta^e_L&=&\left(-\frac{1}{2}+\frac{1}{2}\sin^2(\theta_W)\right)\cos(2\beta)M_Z^2\,,~~~~~\Delta^e_R=-\sin^2(\theta_W)\cos(2\beta)M_Z^2\,.
\end{eqnarray}
The contributions $T^f_{L,R}$ to the running soft masses are usually
ignored, and it is common practice to set them to zero in a numerical scan. 
In our study, we will therefore not consider them any further.

The off-diagonal entries in the soft scalar masses which are induced by
the running are of the same order in $\lambda$ as the high scale ones, with an
additional suppression by~$\eta$. Only for the $LL$ masses of the
down-squarks and charged sleptons, the contributions due to $R_q$ and
$R_l^{(')}$ can be relatively large as those factors take values up to $\sim
35$ in a numerical scan. Generally, however, the main effect of the RG
evolution on the scalar masses is the change of the diagonal elements. 
The masses of the first two generations of $(\tilde{m}_u^2)_{{LL}_{\text{low}}}$,
$(\tilde{m}_u^2)_{{RR}_{\text{low}}}$, $(\tilde{m}_d^2)_{{LL}_{\text{low}}}$
and all three generations of $(\tilde{m}_d^2)_{{RR}_{\text{low}}}$,
$(\tilde{m}_e^2)_{{RR}_{\text{low}}}$ are increased at low energy scales due
to the second terms in Eqs.~(\ref{muLLLow}-\ref{meRRLow}). 
The (33) elements of $(\tilde{m}_u^2)_{{LL}_{\text{low}}}$,
$(\tilde{m}_u^2)_{{RR}_{{Low}}}$ and $(\tilde{m}_d^2)_{{LL}_{\text{low}}}$ can
still remain relatively light, as they also feel the effect of~$R_q$, defined
in Eq.~\eqref{Rq_x}, entering with a negative sign. 
Similarly, the enhancement of all three diagonal entries of 
$(\tilde{m}_e^2)_{{LL}_{\text{low}}}$ is reduced due to 
 the term $-2\eta_NR_l$ which encodes seesaw effects.

\subsection{Low energy mass insertion parameters}\label{App:LowMIs}

With these preparations, we can now formulate the mass insertion parameters at
the low energy scale.

\subsubsection*{Up-type quark sector:}

\begin{eqnarray}
\label{eq:uplowLL12}
(\delta^u_{LL})_{12}&=&\frac{1}{(p^u_{L^{1G}})^2}e^{-i\theta^d_2}\,\tilde{b}_{12}\,\lambda^4,\\
(\delta^u_{LL})_{13}&=&\frac{1}{p^u_{L^{1G}}p^u_{L^{3G}}}e^{-i(4\theta^d_2+\theta^d_3)}(1-\eta\,y_t^2)\,\tilde{b}_{13}\,\lambda^6,\\
(\delta^u_{LL})_{23}&=&\frac{1}{p^u_{L^{1G}}p^u_{L^{3G}}}e^{-i(7\theta^d_2+2\theta^d_3)}(1-\eta\,y_t^2)\,\tilde{b}_{23}\,\lambda^5,
\end{eqnarray}
\begin{eqnarray}
(\delta^u_{RR})_{12}&=&\frac{1}{(p^u_{R^{1G}})^2}e^{-i\theta^d_2}\,\tilde{b}_{12}\,\lambda^4,\\
(\delta^u_{RR})_{13}&=&\frac{1}{p^u_{R^{1G}}p^u_{R^{3G}}}(1-2\eta\,y_t^2)\,\tilde{b}_{13}\,\lambda^6,\\
(\delta^u_{RR})_{23}&=&\frac{1}{p^u_{R^{1G}}p^u_{R^{3G}}}e^{i(5\theta^d_2+\theta^d_3)}(1-2\eta\,y_t^2)\,\tilde{b}_{23}\,\lambda^5,
\end{eqnarray}
\begin{eqnarray}
(\delta^u_{LR})_{11}&=&\frac{\alpha_0\,\upsilon_u}{m_0\,p^u_{L^{1G}}\,p^u_{R^{1G}}}y_u
(1+R^y_u)\left(\frac{\tilde{a}^u_{11}}{y_u}-\frac{\mu(1+R_\mu)}{A_0\,t_\beta}-2\frac{R^a_u}{1+R^y_u}\right)\lambda^8,\\
(\delta^u_{LR})_{22}&=&\frac{\alpha_0\,\upsilon_u}{m_0\,p^u_{L^{1G}}\,p^u_{R^{1G}}}y_c
(1+R^y_u)\left(\frac{\tilde{a}^u_{22}}{y_c}-\frac{\mu(1+R_\mu)}{A_0\,t_\beta}-2\frac{R^a_u}{1+R^y_u}\right)\lambda^4,\\
(\delta^u_{LR})_{33}&=&\frac{\alpha_0\,\upsilon_u}{m_0\,p^u_{L^{3G}}\,p^u_{R^{3G}}}y_t
(1+R^y_t)\left(\frac{\tilde{a}^u_{33}}{y_t}-\frac{\mu(1+R_\mu)}{A_0\,t_\beta}-2\frac{R^a_t}{1+R^y_t}\right),
\end{eqnarray}
\begin{eqnarray}
(\delta^u_{LR})_{12}&=&(\delta^u_{LR})_{21}=(\delta^u_{LR})_{31}\label{deltauLR12Low}=0,\\
(\delta^u_{LR})_{13}&=&-\frac{\alpha_0\,\upsilon_u}{m_0\,p^u_{L^{1G}}\,p^u_{R^{3G}}}\tilde{x}_2\,y_b\,y_t
\left(\frac{\tilde{x}^a_2}{\tilde{x}_2}e^{i(\theta^{\tilde{x}_a}_2-\theta^{\tilde{x}}_2)}+\frac{R^a_t}{1+R^y_t}\right)2\eta\lambda^7,\label{deltauLR13Low}
\end{eqnarray}

\begin{eqnarray}
\label{deltauLR23Low}
(\delta^u_{LR})_{23}&=&\frac{\alpha_0\,\upsilon_u}{m_0\,p^u_{L^{1G}}\,p^u_{R^{3G}}}\Bigg\{-y_s\,y_b\,y_t
\left(\frac{a_s}{y_s}e^{i(\theta^a_s-\theta^y_s)}+\frac{R^a_t}{1+R^y_t}\right)2\eta\lambda^6+\\
\nonumber&+&\lambda^7\Bigg[e^{i\theta^d_2}\tilde{a}^u_{23}(1+R^y_t-\eta\,y_t^2)+2\eta\,y_b\,y_t\Bigg(
e^{i\theta^d_2}\tilde{a}^d_{12}+\left(\frac{a_s}{y_s}e^{i(\theta^a_s-\theta^y_s)}+\frac{R^a_t}{1+R^y_t}\right)\times\\
\nonumber&\times&(\tilde{x}_2\cos(\theta^d_2)-z^d_4\cos(4\theta^d_2+\theta^d_3))+z^d_4e^{i(4\theta^d_2+\theta^d_3)}
\left(e^{i(\theta^a_s-\theta^y_s)}-\frac{z^{d_a}_4}{z^d_4}e^{i(\theta^{z_{d_a}}_4-\theta^{z_d}_4)}\right)\Bigg)\Bigg]\Bigg\},\\
\label{eq:uplowLR32}
 (\delta^u_{LR})_{32}&=&\frac{\alpha_0\,\upsilon_u}{m_0\,p^u_{L^{3G}}\,p^u_{R^{1G}}}(1+R^y_t-2\eta\,y_t^2)e^{i(3\theta^d_2+\theta^d_3)}\tilde{a}^u_{23}\,\lambda^7,
\end{eqnarray}
where, in Eq.~\eqref {deltauLR23Low}, $z^d_4$ and $z^{d_a}_4$ parameterise the
$\mathcal O(\lambda^5)$ NLO corrections of the (22) and (23) elements of the
down-type Yukawa and soft 
trilinear structures, respectively. Originating from the second term of
Eq.~\eqref{YdOperators}, 
$z^d_4 e^{i \theta^{z_d}_4} = y^d_2\, \tilde \delta^d_{3,2_{(4)}} \phi^d_2$, 
so that $\theta^{z_d}_4=6\theta^d_2+4\theta^d_3$.
We see that the term proportional to $\eta\,\lambda^6$, which was generated in
$\tilde{A}^u_{{\text{low}}_{23}}$ via th RG evolution, is the source of the
associated term in $(\delta^u_{LR})_{23}$, which was of order $\lambda^7$ at
the GUT scale.  In Eqs.~(\ref{eq:uplowLL12}-\ref{eq:uplowLR32}) we have
defined the factors
\begin{eqnarray}
\nonumber p^u_{L^{1G}}&=&\sqrt{b_{01}+6.5\,x},~~~~~~~\,
p^u_{L^{3G}}=\sqrt{b_{02}+6.5\,x-2\eta R_q+\frac{\upsilon_u^2}{m_0^2}y_t^2(1+R^y_t)^2}\,,\\
 p^u_{R^{1G}}&=&\sqrt{b_{01}+6.15\,x},~~~~~~p^u_{R^{3G}}=\sqrt{b_{02}+6.15\,x-4\eta R_q+\frac{\upsilon_u^2}{m^2_0}y_t^2(1+R^y_t)^2}\,,~~~~~~\label{pus}
\end{eqnarray}
which are related to the full sfermion mass matrices by
\begin{eqnarray}
\nonumber m_{\tilde{u}_{LL}}&\approx&m_{\tilde{c}_{LL}}\approx m_0\,p^u_{L^{1G}}\,,~~~~\,~~m_{\tilde{t}_{LL}}\approx m_0\,p^u_{L^{3G}}\,,\\
m_{\tilde{u}_{RR}}&\approx&m_{\tilde{c}_{RR}}\approx m_0\,p^u_{R^{1G}}\,,~~~~~~m_{\tilde{t}_{RR}}\approx m_0\,p^u_{R^{3G}}\,,\label{uLR}
\end{eqnarray}
whose GUT scale definitions are given in Eq.~\eqref{fullmasses}.
The $\mu$ parameter at the low energy scale can be estimated by
\begin{eqnarray}
\mu_{\text{low}}\approx\mu\left(1+R_{\mu}\right),~~~~~~R_\mu=4\eta\left(0.9\,g_U^2-\frac{3}{4}y_t^2\right)-3\eta_N\,y_D^2\,.
\end{eqnarray}

\subsubsection*{Down-type quark sector:}

\begin{eqnarray}
(\delta^d_{LL})_{12}&=&\frac{1}{(p^d_{L^{1G}})^2}\tilde{B}_{12}\,\lambda^3,\\
(\delta^d_{LL})_{13}&=&\frac{1}{p^d_{L^{1G}}p^d_{L^{13}}}e^{i\theta^d_2}\frac{\tilde{x}_2^2}{y_b\,y_s}\left(b_{01}-b_{02}+2\eta\,R_q\right)\left(1+\frac{\eta\,y_t^2}{1+R^y_b}\right)\,\lambda^4,\label{dLL13Low}\\
(\delta^d_{LL})_{23}&=&\frac{1}{p^d_{L^{1G}}p^d_{L^{13}}}\frac{y_s}{y_b}\left(b_{01}-b_{02}+2\eta\,R_q\right)\left(1+\frac{\eta\,y_t^2}{1+R^y_b}\right)\,\lambda^2,\label{dLL23Low}
\end{eqnarray}
\begin{eqnarray}
(\delta^d_{RR})_{12}&=&-(\delta^d_{RR})_{13}=\frac{1}{(p^d_{R})^2}e^{i\theta^d_2}\,\tilde{R}_{12}\,\lambda^4,\label{dRR13Low}\\
(\delta^d_{RR})_{23}&=&-\frac{1}{(p^d_{R})^2}\tilde{R}_{12}\,\lambda^4,\label{dRR23Low}
\end{eqnarray}
\begin{eqnarray}
(\delta^d_{LR})_{11}&=&\frac{\alpha_0\,\upsilon_d}{m_0\,p^d_{L^{1G}}\,p^d_{R}}\frac{\tilde{x}_2^2}{y_s}
(1+R^y_d)\left(\frac{\tilde{a}^d_{11}}{\tilde{x}_2^2/y_s}-\frac{\mu\,t_\beta(1+R_\mu)}{A_0}-2\frac{R^a_d}{1+R^y_d}\right)\lambda^6,\\
(\delta^d_{LR})_{22}&=&\frac{\alpha_0\,\upsilon_d}{m_0\,p^d_{L^{1G}}\,p^d_{R}}y_s
(1+R^y_d)\left(\frac{\tilde{a}^d_{22}}{y_s}-\frac{\mu\,t_\beta(1+R_\mu)}{A_0}-2\frac{R^a_d}{1+R^y_d}\right)\lambda^4,\\
(\delta^d_{LR})_{33}&=&\frac{\alpha_0\,\upsilon_d}{m_0\,p^d_{L^{3G}}\,p^d_{R}}y_b
(1+R^y_b)\left(\frac{\tilde{a}^d_{33}}{y_b}-\frac{\mu\,t_\beta(1+R_\mu)}{A_0}-2\frac{R^a_b}{1+R^y_b}\right)\lambda^2,
\end{eqnarray}
\begin{eqnarray}
(\delta^d_{LR})_{12}&=&-(\delta^d_{LR})_{21}=(\delta^d_{LR})_{13}=\frac{\alpha_0\,\upsilon_d}{m_0\,p^d_{L^{1G}}\,p^d_{R}}(1+R^y_d)\tilde{a}^d_{12}\,\lambda^5,\\
(\delta^d_{LR})_{23}&=&\frac{\alpha_0\,\upsilon_d}{m_0\,p^d_{L^{1G}}\,p^d_{R}}y_s(1+R^y_d)
\left(\frac{\tilde{a}^d_{23}}{y_s}+2\frac{\eta\,y_t^2}{1+R^y_b}\left(\frac{a_t}{y_t}+\frac{R^a_d}{1+R^y_d}\right)\right)\lambda^4,~~~~~~~~~~\label{dLR23Low}\\
(\delta^d_{LR})_{31}&=&\frac{\alpha_0\,\upsilon_d}{m_0\,p^d_{L^{3G}}\,p^d_{R}}e^{-i\theta^d_2}(1+R^y_b)\tilde{a}^d_{31}\,\lambda^6,
\end{eqnarray}

\begin{eqnarray}
\nonumber\label{dLR32Low}(\delta^d_{LR})_{32}&=&\frac{\alpha_0\,\upsilon_d}{m_0\,p^d_{L^{3G}}\,p^d_{R}}(1+R^y_b)y_b\Bigg(\frac{\tilde{a}^d_{32}}{y_b}+2\eta y_t^2\frac{y_s^2}{y_b^2}\Bigg[\frac{2(1+R^y_b)+\eta y_t^2}{2(1+R^y_b)^2}\frac{\tilde{a}^d_{23}}{y_s}\\
&+&\left(\frac{a_t}{y_t}+\frac{R^a_d}{1+R^y_d}\right)\frac{(1+R^y_d)^2}{(1+R^y_b)^3}\Bigg]\Bigg)\lambda^6,
\end{eqnarray}
where
\begin{eqnarray}
p^d_{L^{1G}}&=&\sqrt{b_{01}+6.5\,x},~~~~~p^d_{L^{3G}}=\sqrt{b_{02}+6.5\,x-4\eta R_q},~~~~~p^d_R=\sqrt{1+6.1x},\label{pds}~~~~~~~~
\end{eqnarray}
such that
\begin{eqnarray}
\nonumber m_{\tilde{d}_{LL}}&\approx&m_{\tilde{s}_{LL}}\approx m_0\,p^d_{L^{1G}}\,,~~~~~~m_{\tilde{b}_{LL}}\approx m_0\,p^d_{L^{3G}}\,,\\
m_{\tilde{d}_{RR}}&\approx&m_{\tilde{s}_{RR}}\approx m_{\tilde{b}_{RR}}\approx m_0\,p^d_{R}\,.\label{dLR}
\end{eqnarray}

\subsubsection*{Charged lepton sector:}

\begin{eqnarray}
 (\delta^e_{LL})_{12}&=&-(\delta^e_{LL})_{23}=\frac{1}{(p^e_{L})^2}\left(\tilde{R}_{12}-2\eta_N\tilde{E}_{12}\right)\lambda^4,\label{eLL12Low}\\
 (\delta^e_{LL})_{13}&=&-\frac{1}{(p^e_{L})^2}\left(\tilde{R}_{12}-2\eta_N\tilde{E}^*_{12}\right)\lambda^4,
\end{eqnarray}
\begin{eqnarray}
 (\delta^e_{RR})_{12}&=&-\frac{1}{(p^e_{R^{1G}})^2}e^{i\theta^d_2}\frac{\tilde{B}_{12}}{3}\,\lambda^3,\\
(\delta^e_{RR})_{13}&=&\frac{1}{p^e_{R^{1G}}\,p^e_{R^{3G}}}\frac{\tilde{B}_{13}}{3}\,\lambda^4,\label{eRR13Low}\\
(\delta^e_{RR})_{23}&=&\frac{1}{p^e_{R^{1G}}\,p^e_{R^{3G}}}3\tilde{B}_{23}\,\lambda^2,
\end{eqnarray}
\begin{eqnarray}
(\delta^e_{LR})_{11}&=&\frac{1}{p^e_{L}\,p^e_{R^{1G}}}\frac{\upsilon_d\,\alpha_0}{m_0}
\frac{\tilde{x}_2^2}{3\,y_s}(1+R^y_e)\left(\frac{y_s}{\tilde{x}_2^2}\tilde{a}^d_{11}-\frac{\mu\,t_\beta}{A_0}(1+R_\mu)-2\frac{R^a_e}{1+R^y_e}\right)\lambda^6,~~~~~~\\
 (\delta^e_{LR})_{22}&=&\frac{1}{p^e_{L}\,p^e_{R^{1G}}}\frac{\upsilon_d\,\alpha_0}{m_0}
3\,y_s(1+R^y_e)\left(\frac{\tilde{a}^d_{22}}{y_s}-\frac{\mu\,t_\beta}{A_0}(1+R_\mu)-2\frac{R^a_e}{1+R^y_e}\right)\lambda^4,\\
 (\delta^e_{LR})_{33}&=&\frac{1}{p^e_{L}\,p^e_{R^{3G}}}\frac{\upsilon_d\,\alpha_0}{m_0}
y_b(1+R^y_e)\left(\frac{\tilde{a}^d_{33}}{y_b}-\frac{\mu\,t_\beta}{A_0}(1+R_\mu)-2\frac{R^a_e}{1+R^y_e}\right)\lambda^2,
\end{eqnarray}
\begin{eqnarray}
 (\delta^e_{LR})_{12}&=&\frac{1}{p^e_{L}\,p^e_{R^{1G}}}\frac{\upsilon_d\,\alpha_0}{m_0}(1+R^y_e)e^{i\theta^d_2}\tilde{a}^d_{12}\,\lambda^5,\\
 (\delta^e_{LR})_{13}&=&\frac{1}{p^e_{L}\,p^e_{R^{3G}}}\frac{\upsilon_d\,\alpha_0}{m_0}\left((1+R^y_e)\tilde{a}^d_{31}
+2
\eta_N\,y_D\,R_\nu\,y_b\left(\frac{\alpha_D}{y_D}+\frac{R^a_e}{1+R^y_e}\right)\right)\lambda^6,~~~~~~~~\\
(\delta^e_{LR})_{21}&=&(\delta^e_{LR})_{31}=-\frac{1}{p^e_{L}\,p^e_{R^{1G}}}\frac{\upsilon_d\,\alpha_0}{m_0}(1+R^y_e)e^{-i\theta^d_2}\tilde{a}^d_{12}\,\lambda^5,\\
 (\delta^e_{LR})_{23}&=&\frac{1}{p^e_{L}\,p^e_{R^{3G}}}\frac{\upsilon_d\,\alpha_0}{m_0}\left((1+R^y_e)\tilde{a}^e_{23}
+2 \eta_N\,y_D\,R_\nu\,y_b\left(\frac{R^a_\nu}{R_\nu}+\frac{R^a_e}{1+R^y_e}\right)\right)\lambda^6,\\
 (\delta^e_{LR})_{32}&=&\frac{1}{p^e_{L}\,p^e_{R^{1G}}}\frac{\upsilon_d\,\alpha_0}{m_0}(1+R^y_e)3\,\tilde{a}^d_{23}\,\lambda^4,
\end{eqnarray}
where
\begin{eqnarray}
 p^e_L&=&\sqrt{1+0.5\,x-2\eta_N\,R_l},~~~~p^e_{R^{1G}}=\sqrt{b_{01}+0.15\,x},~~~~p^e_{R^{3G}}=\sqrt{b_{02}+0.15\,x},~~~~~~~~
\end{eqnarray}
such that
\begin{eqnarray}
\nonumber m_{\tilde{e}_{LL}}&\approx&m_{\tilde{\mu}_{LL}}\approx m_{\tilde{\tau}_{LL}}\approx m_0\,p^e_{L}\,,\\
m_{\tilde{e}_{RR}}&\approx&m_{\tilde{\mu}_{RR}}\approx  m_0\,p^e_{R^{1G}}\,,~~~~~~m_{\tilde{\tau}_{RR}}\approx m_0\,p^e_{R^{3G}}\,.\label{eLR}
\end{eqnarray}

\end{appendix}






\end{document}